\documentclass[11pt]{article}
\usepackage{hyperref}
\usepackage{amsthm}
\usepackage{amsmath}
\usepackage{amsfonts}
\usepackage{amssymb}
\usepackage[ruled, vlined]{algorithm2e}
\usepackage{graphicx}
\usepackage{natbib}
\graphicspath{{figures/}}

\hypersetup{
    colorlinks=true,
    linkcolor=blue,
    urlcolor=black,
    citecolor=blue
    }

\title{\vspace{-1cm}
Contour Location for Reliability in Airfoil Simulation Experiments \\
using Deep Gaussian Processes}
\author{Annie S. Booth\thanks{Corresponding author: Department of Statistics, 
  NC State University, {\tt annie\_booth@ncsu.edu}} 
  \and S. Ashwin Renganathan\thanks{Department of Aerospace Engineering,
  Penn State} 
  \and Robert B. Gramacy\thanks{Department of Statistics, Virginia Tech}}
\date{\today}

\begin{document}

\maketitle

\begin{abstract}
Bayesian deep Gaussian processes (DGPs) outperform ordinary GPs as surrogate
models of complex computer experiments when response surface dynamics are
non-stationary, which is especially prevalent in aerospace simulations.  Yet
DGP surrogates have not been deployed for the canonical downstream task in
that setting: reliability analysis through contour location (CL).  
In that context, we are motivated by a simulation of an RAE-2822 transonic airfoil which
demarcates efficient and inefficient flight conditions.
Level sets separating passable versus failable 
operating conditions are best learned through
strategic sequential designs.  There are two limitations to modern CL
methodology which hinder DGP integration in this setting.  First,
derivative-based optimization underlying acquisition functions is thwarted by
sampling-based Bayesian (i.e., MCMC) inference, which is essential for DGP
posterior integration. Second, canonical acquisition criteria, such as
entropy, are famously myopic to the extent that optimization may even be
undesirable. Here we tackle both of these limitations at once, proposing a
hybrid criterion that explores along the Pareto front of entropy and
(predictive) uncertainty, requiring evaluation only at strategically located
``triangulation'' candidates. We showcase DGP CL performance in several
synthetic benchmark exercises and on the RAE-2822 airfoil.
\end{abstract}


\section{Introduction} \label{sec:intro}

Computer simulations are increasingly utilized for experimentation when direct
manipulation is impossible or infeasible.  Although diverse applications
abound \citep[e.g.,][]{tietze2015model,bremer2019globus,lippe2019using}, here
we are motivated by one in particular: physics-based simulations in
aeronautics. These involve computational fluid dynamics (CFD\footnote{See
Supplementary Material A for a complete list of acronyms.}) simulation
requiring the numerical solution of a set of stiffly coupled nonlinear partial
differential equations satisfying conservation laws
\citep[e.g.,][]{pamadi2004aerodynamic,vassberg2008development,mehta2014modeling}.
A specific example is aerodynamic modeling of the interaction between an
aircraft, aircraft wing, or wing section (a.k.a.,``airfoils'') and the
surrounding airflow at transonic speeds (i.e., flight at $\sim 70-80 \%$ of
the speed of sound). Transonic flow straddles the line between subsonic
(flying slower than the speed of sound) and supersonic (faster than sound)
flow that makes it particularly complex to model. When the computational costs
of obtaining runs severely limits how many may be collected, a ``surrogate
model'' or ``emulator'', which is a fitted statistical model trained on data
from a computer experiment, may be essential for downstream tasks requiring
model evaluation at unobserved inputs.

The canonical surrogate is a Gaussian process
\citep[GP;][]{santner2018design,gramacy2020surrogates}, but in their
traditional form, GPs suffer from the assumption of stationarity.  They must
impart similar dynamics across the entire input space which limits their
ability to cope with regime shifts, as are common in aerospace simulations.
 Aircraft experience ``shocks'' when the sound barrier is
broken. Several adaptations have been considered to allow for non-stationary
GP flexibility \citep[see][for a review]{sauer2023non}, but there are
drawbacks. Non-stationary kernels suit spatial applications with low input
dimension \citep{higdon1999non,paciorek2003nonstationary}. Partition- or
treed-GPs divy up the input space
\citep{gramacy2008bayesian,bitzer2023hierarchical}, sacrificing global scope.
Locally approximate GPs are data hungry \citep{gramacy2015local}.

The deep Gaussian process \citep[DGP;][]{damianou2013deep} shares many of these
same ingredients, but has fewer drawbacks. DGPs use stationary GPs to 
spatially warp the original inputs. 
Warped inputs were first proposed in the spatial statistics community
\citep{sampson1992nonparametric,schmidt2003bayesian}, but have recently been
popularized by machine learners who showcased DGP prowess on large-scale
classification and regression tasks \citep{damianou2013deep,dunlop2018how}.
The challenge in DGP inference lies in learning high-dimensional
latent warping variables.  With large training data sizes and a penchant
for computational thrift, many in machine learning embraced approximate
variational inference \citep[VI;][]{bui2016deep,salimbeni2017doubly}.

Yet the supposed thriftiness of approximate VI often disappoints because the
algorithms require careful tuning.  Moreover, optimization in lieu of full
posterior integration sacrifices uncertainty quantification (UQ), which is
essential to safety/reliability tasks, especially in data-poor settings. 
We instead embrace the fully-Bayesian inferential scheme
of \citet{sauer2023active}, leveraging modern advances in Markov chain Monte
Carlo (MCMC). This fully-Bayesian DGP has been shown to outperform VI-based
approximate DGP competitors \citep{sauer2022vecchia,sauer2023non} and to excel
in surrogate modeling settings where training data is limited
\citep{sauer2023active}.

\subsection{RAE-2822 airfoil computer experiment}
\label{sec:motive}

\begin{figure}[ht!]
\centering
  \begin{minipage}{6.5cm}
  \includegraphics[width=5.5cm]{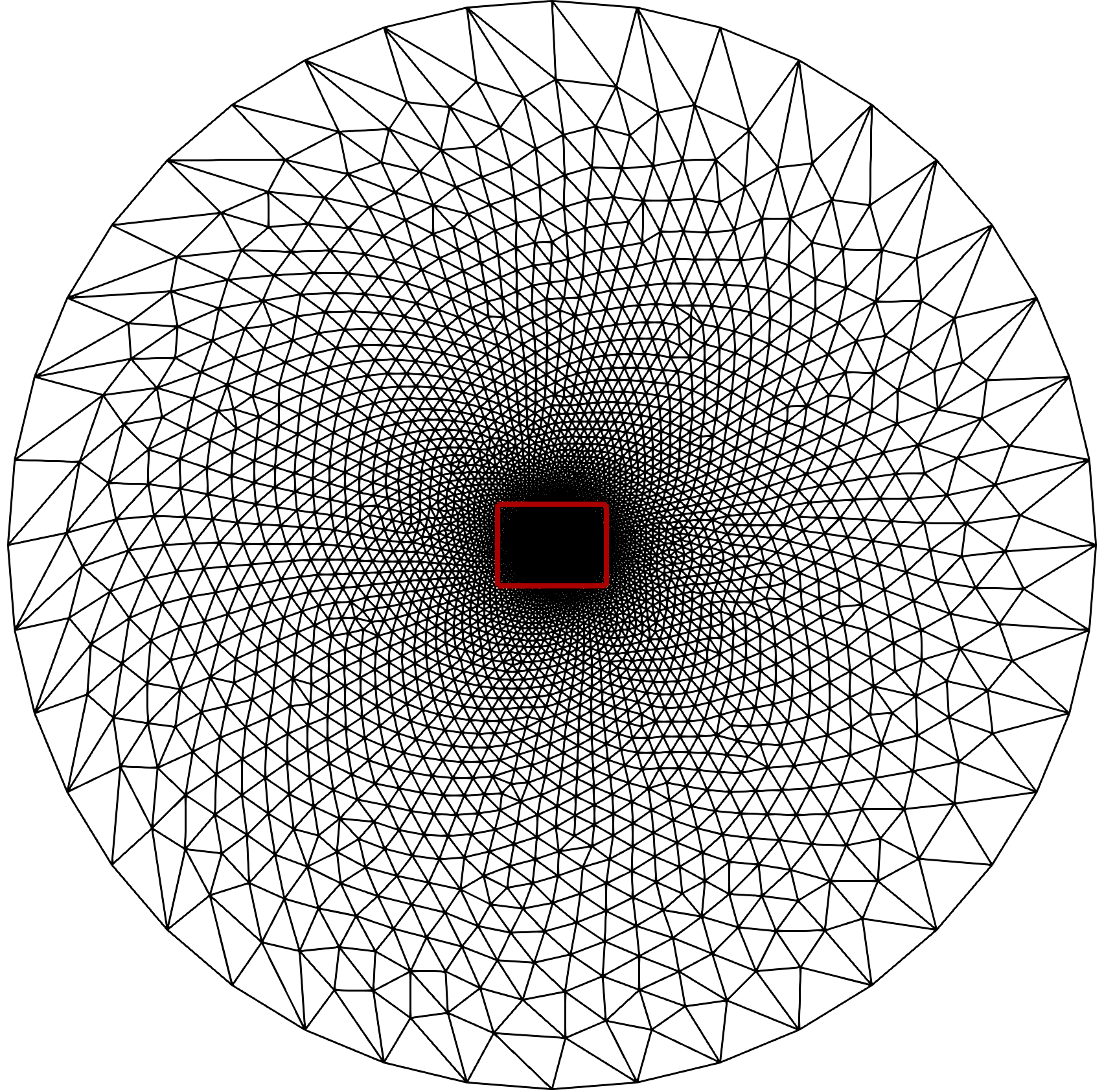}
  \end{minipage}
  \begin{minipage}{7cm}
  \includegraphics[width=6.5cm]{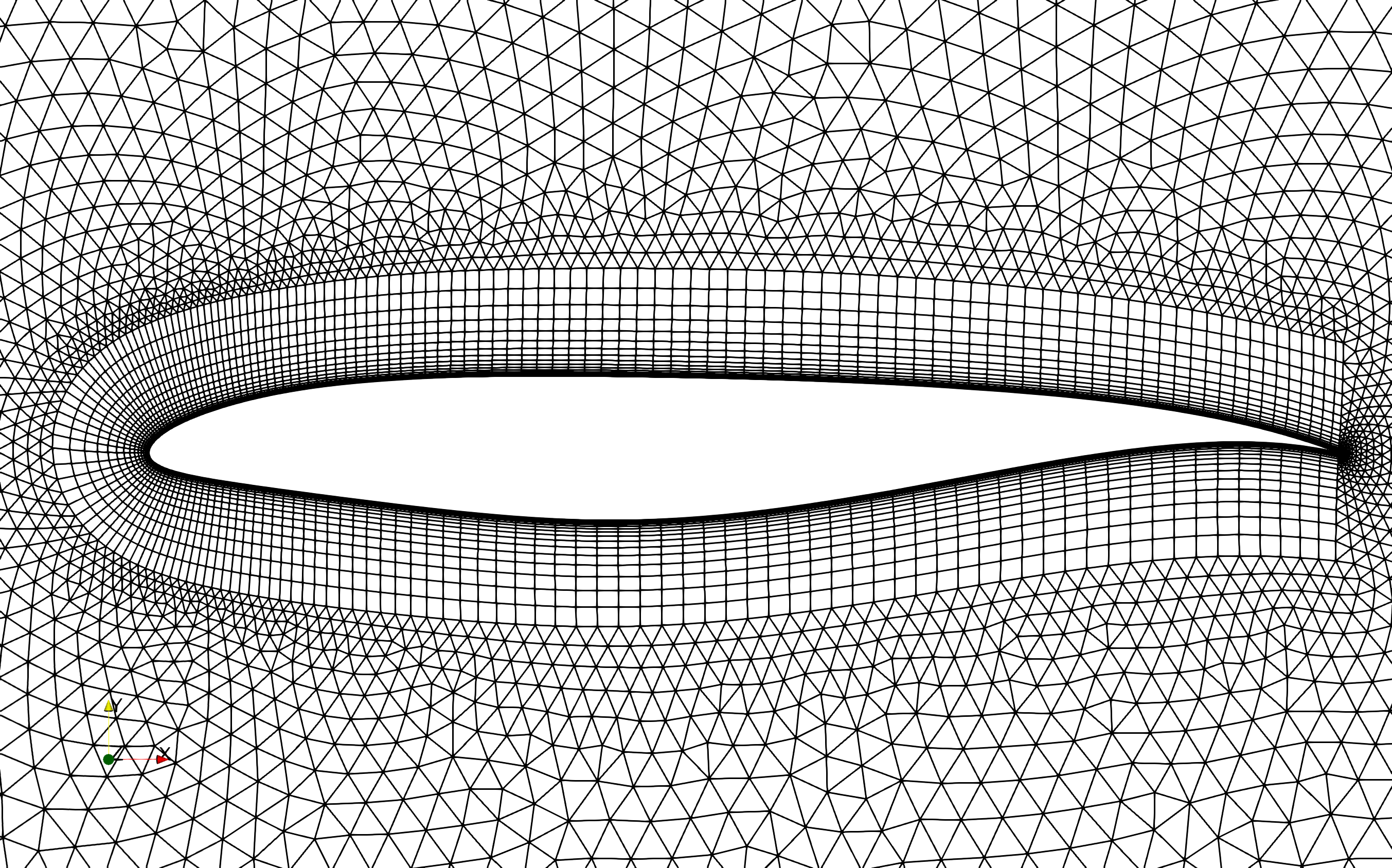}
  \end{minipage}
  \caption{Computational mesh. {\it Left:} Full domain showing all 
  $27,857$ mesh elements.  {\it Right:} Near-field
  mesh refinement for the RAE-2822 airfoil to capture the boundary layer and
  shock.}
  \label{fig:airfoil_grid}
\end{figure}

Here we are motivated by a computer simulation of transonic flow past 
an RAE-2822 airfoil.  The model, utilizing Reynolds Averaged
Navier-Stokes equations, is solved via SU2 ~\citep{economon2016su2}, a
public simulation suite.  
A spherical domain of $100$ airfoil chord length radius is used to model the 
fluid domain, which is discretized with 27,857 triangular and 
hexahedral mesh elements.  The hexahedral elements are employed 
closer to the surface of the airfoil to capture the boundary 
layer.  Additionally, the mesh density is made deliberately higher 
in the near-field of the airfoil to resolve the shock better; see 
Figure \ref{fig:airfoil_grid}.  Conservation laws (mass, momentum, and 
energy) are enforced in each mesh element forming a system of nonlinear 
equations which are solved iteratively.  The computational cost of the 
solution scales cubically with the number of mesh elements. 
All simulations seek the 
steady-state solution, with pseudo time-stepping.  In serial execution, 
each simulation takes 20-30 minutes of wallclock time for convergence, and 
about 150 seconds with 8 parallel MPI processes.

We consider seven input variables: three freestream conditions and four shape
parameters.  Freestream conditions include angle of attack with bounds $[0,
10]$, Mach number ranging in $[0.7, 0.9]$, and Reynolds number with bounds
$[5\times 10^6, 15\times 10^6]$. The airfoil shape and surrounding mesh
are modified by displacing free form deformation control points around the airfoil.  
The quantity of interest is the
lift-drag ratio $L/D \equiv C_l / C_d$, where $C_l$ and $C_d$ are the lift and
drag coefficients, respectively. $L/D$ is a measure of the aerodynamic
efficiency of an aircraft and directly impacts the amount of fuel required.
Strict airworthiness standards from the Federal Aviation Administration limit
the amount of fuel an aircraft can burn per passenger mile, since more jet
fuel burned results in more greenhouse gas emissions, thus aircraft
designers aim to meet a minimum required $L/D$. Our current goal is to
identify where $L/D < 3$, which is considered failure.

\begin{figure}[ht!]
\centering
\includegraphics[width=17.5cm,trim=10 250 10 100]{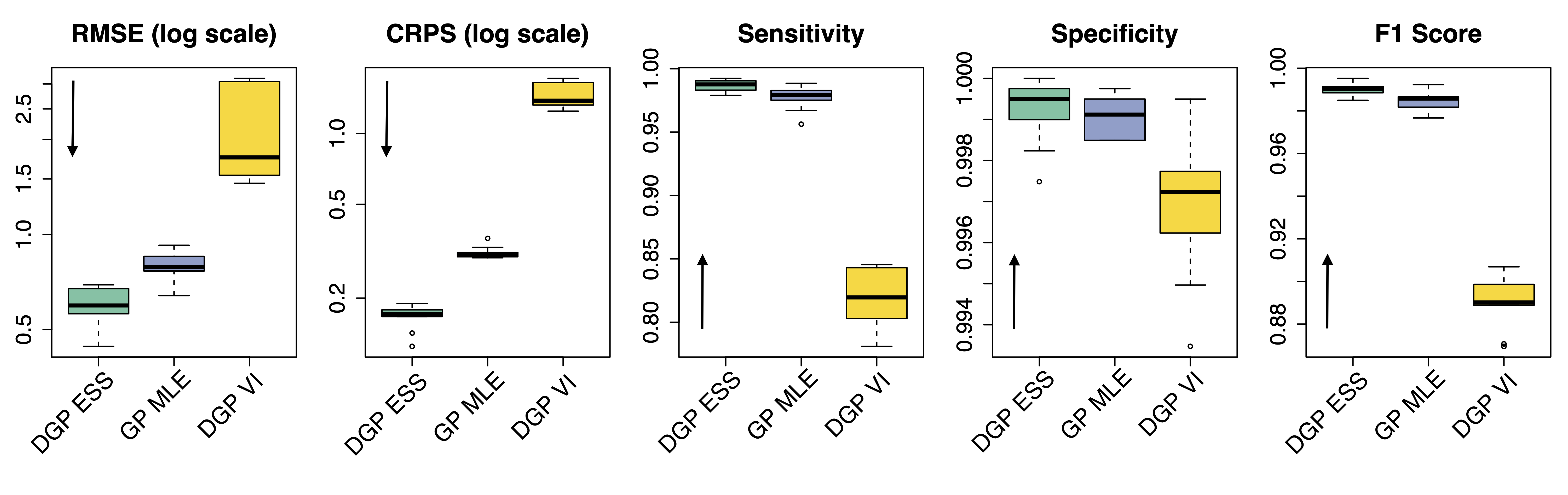}
\caption{Out-of-sample prediction metrics for surrogate fits to static 
LHS designs of size $n = 500$ from the RAE-2822 simulation.  Failures defined at 
$L/D < 3$.  Arrows along $y$-axes indicate which direction is preferred.
Boxplots show 10 MC repetitions.}
\label{fig:su2static}
\end{figure}

DGPs are relatively new to this area, but expert knowledge suggests there
is non-stationarity in the response surfaces of these simulations.  To test
this, we fit both shallow and deep GP surrogates to simulations obtained from
a 500-run space-filling Latin hypercube sample \citep[LHS;][]{mckay1979comparison}
in the 7d input space, specifically: the Bayesian
DGP using elliptical slice sampling (DGP ESS) of \citet{sauer2023active}; 
a traditional stationary GP via maximum likelihood estimates (GP MLE); and 
the approximate VI DGP (DGP VI) of \citet{salimbeni2017doubly}.
Further details about these comparators will be revealed as the paper progresses.
Predictive performance is assessed on a hold-out 4,500-point LHS
via root mean squared
error (RMSE, measuring accuracy) and continuous ranked probability score
\citep[CRPS, measuring UQ;][]{gneiting2007strictly}.  See the left panels of Figure
\ref{fig:su2static}.  Lower is better for both (Supp.~B).

The Bayesian DGP is both more accurate and offers better UQ, but
improved prediction is just the tip of an iceberg. Surrogate models are most
often a means to another end; they are used for downstream tasks such as
Bayesian optimization \citep[BO; e.g.,][]{jones1998efficient}, active learning
\citep[AL; e.g.,][]{cohn1994neural}, and calibration
\citep[e.g.,][]{kennedy2001bayesian}.  Here we are interested in contour
location (CL), a common task in aeronautic reliability analysis
\citep[e.g.,][]{stanford2022gradient,renganathan2022multifidelity,renganathan2023camera}.  
In CL, the
surrogate's task is to identify a specific level set in the response surface,
one that separates system ``passes'' from ``failures''.  A surrogate must accurately locate
failure regions in order to provide precise quantification of failure
probabilities.

With our 7d airfoil simulator, we seek to identify 
efficient versus inefficient ($L/D<3$) flight conditions. The right panels of
Figure \ref{fig:su2static} provide out-of-sample classification metrics
measuring sensitivity, specificity, and F1 scores
(Supp.~B) on surrogates fit to RAE-2822 simulations.
Observe that the Bayesian DGP also performs well on these metrics (higher is
better for all).  
Yet there is further scope for improvement;
space-filling designs are, of course, inefficient for finding
lower-dimensional manifolds like level sets.
Furthermore, the computing time required to evaluate the computer
model at these 500 points was nearly 20 hours.
With a strategic design, we seek to obtain superior predictive
performance with fewer evaluations of the expensive model.

\subsection{Sequential design for contour location}

We are interested in finding inputs that separate ``pass'' and ``fail''
regimes under a limited simulation budget. In this context, it makes sense
to build up the design sequentially to strategically hone in on this target,
i.e., CL -- a form of AL.  There are
two keys to effective AL: (a) training a good surrogate with the data 
you already have; and (b) using the surrogate to solve an acquisition 
criterion to determine the next run of the simulator which 
will be used to augment the data. These steps are repeated until the 
simulation budget is exhausted or until some performance metric is 
satisfied.  The (Bayesian) DGP provides (a), but it is less amenable 
to (b) compared to the ordinary GP.

For CL, acquisition criteria target the contour through uncertainty in the
pass/fail prediction, which is quantifiable through surrogate posterior
predictions.  The most popular criterion, based on entropy
\citep{gray2011entropy}, converts classification probabilities into a metric
which is maximized for 50/50 probabilities, where classification
uncertainty is highest.  But entropy and its variants have two downsides.  One
is that finding the input location with highest entropy requires a multi-start
numerical optimizer in order to circumvent disparate regions (local
optima) of high entropy, separated by wide gulfs of numerically zero entropy.
This difficulty is compounded when there are disconnected level sets
separating regimes. MCMC-based inferential schemes make things more
challenging still: the surrogate (i.e., a DGP) doesn't provide a single
predictive surface but thousands, precluding library-based optimization.
Second, and perhaps more importantly, the level set is a continuum, not a
singleton as in BO and other AL variants.  Although entropy does indeed target
the contour, it is famously myopic \citep[e.g.,][]{cole2022entropy} as an
acquisition function: it targets the most probable boundary crossing point,
not exploration of the entire manifold. The result is clumped acquisitions.

These drawbacks have inspired a cottage industry of workarounds.  One simple
solution involves evaluating entropy on a discrete set of space-filling
candidates, like an LHS.
This is straightforward, and has some advantages over continuous numerics,
including parallel evaluation and simple modular implementation.  (Evaluating
entropy on candidates is a post-processing step, whereas numerical
optimization must be embedded within the surrogate.) Yet after an initial
handful of acquisitions, most candidates reside in zero-entropy regions,
resulting in essentially random selection and a coarse view of the target
contour.  Alternatively, some have adapted expected improvement ideas from BO
to fit the CL setting, attempting to balance both exploration and exploitation
\citep{ranjan2008sequential,bichon2008efficient,picheny2010adaptive}. Others
have found success within stepwise uncertainty reduction (SUR) BO frameworks,
in which entropy (and other criteria) are numerically integrated over the
input domain \citep{chevalier2014fast,marques2018contour}. However, the
requisite numerical quadrature can be difficult, making SUR-based
strategies computationally daunting in moderate input dimension.
\citet{cole2022entropy} suggested a hybrid-candidate-local numerical
optimization to favor local optima in the entropy surface, a sensible but
ultimately {\em ad hoc} enterprise: deliberately doing worse at one thing
(solving for high entropy) to do better at another (avoiding myopia).
Nevertheless, it out-performs the methods cited above in many exercises.  But
it requires derivatives, making it ill-suited to MCMC.

Here we explore an idea from the BO literature that strategically allocates
candidates between existing design points through the use of Delaunay
triangulation \citep{gramacy2022triangulation}. So-called ``tricands'' are
space-filling in an adaptive way, guaranteeing that there are candidates in
promising regions of the input space for acquisition. Optimization on
surrogate predictive quantities, cumbersome with MCMC-based surrogates, is
replaced with a geometric solver \citep{quickhull,geometry} that doesn't
consult the surrogate at all. As candidates, tricands enjoy parallel
evaluation downstream with any surrogate, even MCMC-based ones.  For BO, they
are superior to LHS candidates {\em and} numerically optimized acquisitions.
For CL, well-chosen candidates provide an opportunity to address multiple
criteria simultaneously, for example to correct the myopia in pure
entropy-based acquisition.  Toward that end, we propose acquisitions on the
``Pareto front'' of tricands evaluations of CL entropy and ordinary predictive
uncertainty.  This is both effective, and enjoys a simple implementation as a
post-processing step on surrogate prediction. While similar to ideas of
\citet{bryan2005active}, who utilize the product of posterior variance and
entropy, our Pareto acquisitions are agnostic to the relative scale
of predictive variance.

\subsection{Roadmap}

Each ingredient in this scheme (DGP, tricands, Pareto front), separately
offers a performance bump over its state-of-the-art analogue (GP,
derivative-based acquisition, entropy-only).  Over the course of this paper,
we will offer intuitive explanations of these improvements and demonstrate
superiority empirically with realistic benchmarking exercises.   We
acknowledge that these ingredients, taken separately, may not comprise a
substantial novel contribution. But their application in the CL context is
novel, both separately and together, and they have never been used in an
airfoil reliability context.  DGPs, whether VI- or MCMC-based, have never been
entertained for CL, and both tricands and Pareto acquisitions are new to CL.
To ease use downstream and support reproducibility, we provide a fully open
source implementation as a supplement to this document, and in public
repositories.\footnote{Code to reproduce all examples and figures may be found
at \url{https://bitbucket.org/gramacylab/deepgp-ex/}. DGP surrogate entropy
calculations are provided as an update to the {\tt deepgp} package on CRAN
\citep{deepgp}.}

The remainder of this paper is laid out as follows.  Section \ref{sec:review}
provides a review of the surrogate-CL state-of-the-art: GPs and entropy-based
acquisition. Section \ref{sec:method} details the main, modern ingredients
-- DGP surrogates, triangulation candidates, and Pareto front acquisitions
-- and how they weave together to form our contribution.  Section 
\ref{sec:results} offers implementation
details and empirical results on synthetic examples.
Section \ref{sec:su2} returns to our motivating airfoil simulation.  
We offer a brief discussion in Section \ref{sec:discuss}.

\section{Stationary GP Contour Location}\label{sec:review}

Here we review the state-of-the-art in GP-based entropy contour location.
The more modern ingredients of our approach (DGPs, tricands, etc.) will be
reviewed as they are adapted to suit CL in Section \ref{sec:method}.

\subsection{Gaussian process surrogates}

Let $f:\mathbb{R}^d\rightarrow \mathbb{R}$ denote a blackbox computer
simulator; $X_n$ be an $n\times d$ matrix of simulation inputs, with
rows $x_i^\top$; and let $y_n = f(X_n)$ represent an $n$-vector of
outputs. A traditional GP prior assumes 
\begin{equation}\label{eq:gpprior}
y_n \sim \mathcal{N}_n\left(\mu, \Sigma(X_n)\right)
\quad\textrm{where}\quad
\Sigma(X_n)^{ij} = \Sigma(x_i, x_j) = 
\tau^2\left(k\left(\sum_{h=1}^d\frac{(x_{ih} - x_{jh})^2}{\theta_h}\right) 
+ \eta\mathbb{I}_{i=j}\right).
\end{equation}
While $\mu$ may be linear in columns of $X_n$, we specify $\mu = 0$ without
loss of generality.  Popular choices of kernel $k(\cdot)$ are the squared
exponential or Mat\`ern \citep{stein1999interpolation}, but any positive
definite function will suffice.  Hyperparameters $\tau^2$, $\theta =
[\theta_1, \dots, \theta_d]$, and $\eta$ govern the scale, lengthscale, and
nugget/noise respectively.  We focus here on deterministic computer
simulations, fixing $\eta = 1\times 10^{-6}$ throughout, to interpolate
observations. Vectorized $\theta$ allows for coordinate-wise anisotropic
modeling. Fixing $\theta = \theta_1 = \dots = \theta_d$ enforces isotropy.
These unknown hyperparameters may be estimated through the likelihood
by maximization or MCMC. See
\citet{santner2018design,gramacy2020surrogates} for review.

\begin{figure}[ht!]
\centering
\includegraphics[width=17.5cm,trim=15 0 10 10]{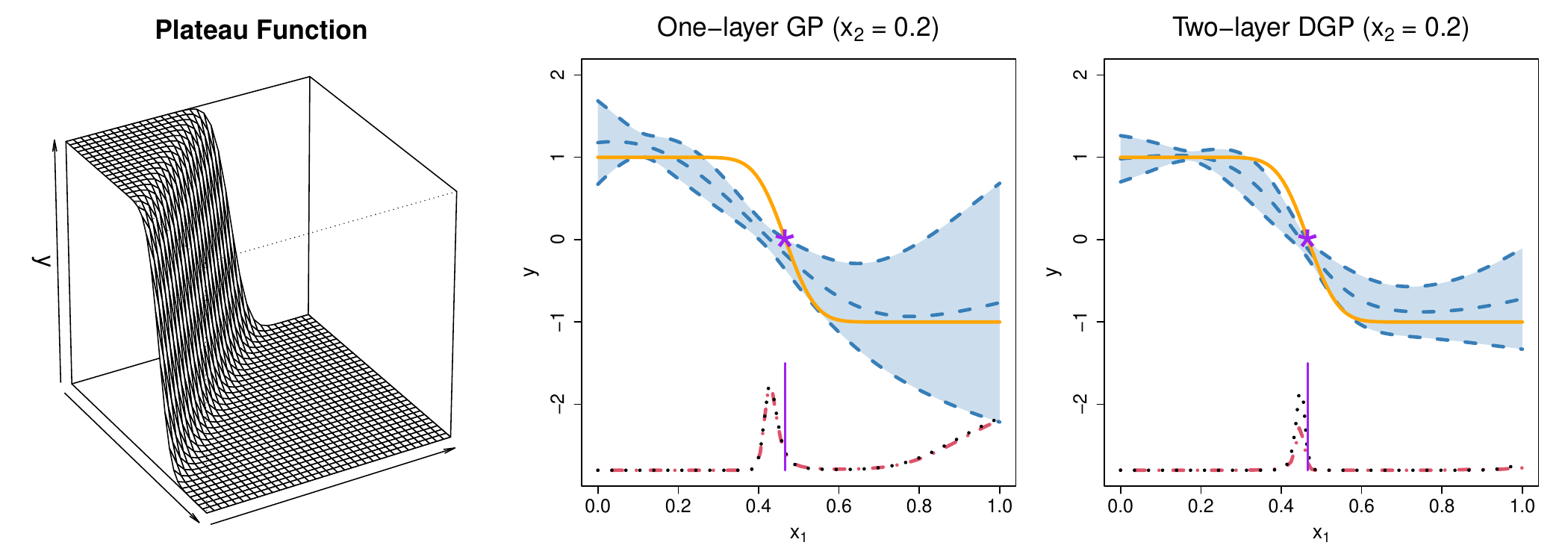} \\
\includegraphics[width=15cm,trim=0 60 0 0]{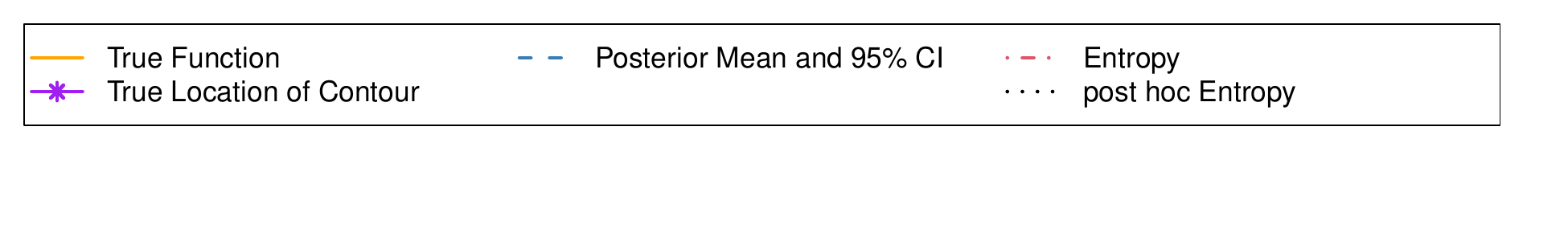}
\caption{{\em Left:} plateau function in 2d.  {\em Middle and right:} 
predictive distribution
along a slice in $x_2$ with entropy calculations for an ordinary (left)
and deep (right) GP.}
\label{fig:oneD}
\end{figure}

Given $\{X_n, y_n\}$, posterior predictions at an $n_p\times d$
matrix of unobserved inputs $\mathcal{X}$ follow
\begin{equation}\label{eq:gppred}
\begin{aligned}
\mathcal{Y} \mid X_n, y_n \sim\mathcal{N}_{n_p}\left(\mu_Y, \Sigma_Y\right)
\quad\textrm{where}\quad
\mu_Y &= \Sigma(\mathcal{X}, X_n)\Sigma(X_n)^{-1} y_n \\
\Sigma_Y &= \Sigma(\mathcal{X}) - \Sigma(\mathcal{X}, X_n)\Sigma(X_n)^{-1}
\Sigma(X_n, \mathcal{X}).
\end{aligned}
\end{equation}
As a visual, consider the 2d ``plateau'' function displayed in the left panel
of Figure \ref{fig:oneD} and featured later in Section \ref{sec:synthetic}. 
We trained a GP surrogate to data obtained from a uniformly random design of size $n =
15$; a 1d slice of the predictive surface (with fixed $x_2 = 0.2$) is
displayed in the center panel.  For now, focus on just the blue lines/shading.
The GP offers an adequate nonlinear fit to the surface, considering the
sparsity of the training data in 2d, but it struggles to balance the disparity
between the flat regions and the steep drop.  This  GP
is restricted by the assumption of {\it stationarity}, a byproduct of
$\Sigma(\cdot)$ being solely a function of relative distances.  A stationary
GP will struggle to balance steep drops with flat regions; it must compromise
between oversmoothing the drop and undersmoothing the flats.  A DGP, previewed
in the right panel, is more flexible in handling this transition. Surrogate
predicted mean and variance estimates are essential components of
AL enterprises like CL.

\subsection{Entropy contour location}

The most popular acquisition criterion for CL is entropy
\citep[e.g.,][]{oakley2004estimating,marques2018contour,cole2022entropy}.  Let
$g$ represent a threshold level such that $f(x) > g$ indicates failure.
Denote $p_x = \mathbb{P}\left(f(x) > g\right)$ as the probability of observing
a failure at location $x$. The entropy criterion
\begin{equation}\label{eq:ent}
H(x) = -p_x\log(p_x)- (1 - p_x)\log(1 - p_x),
\end{equation}
will be high in regions of pass/fail uncertainty (i.e., $p_x \approx 0.5$).
Under a GP surrogate, failure probabilities may be evaluated through a
Gaussian CDF $\Phi$:
\begin{equation}\label{eq:p}
p_x = \mathbb{P}\left(f(x) > g\right) = 
	1 - \Phi\left(\frac{g - \mu_Y(x)}{\sigma_Y(x)}\right),
\end{equation}
where $\mu_Y(x)$ and $\sigma_Y(x) = \sqrt{\Sigma_Y(x)}$ are provided in
Eq.~(\ref{eq:gppred}). Consider again the center panel of
Figure~\ref{fig:oneD}.  Define $g = 0$ such that $y > 0$ represents failure.
The true location of the failure contour (where the true line crosses $g = 0$)
is marked by the purple star and vertical line. The red and black lines along
the $x$-axis show entropy $H(x)$ evaluated along a dense grid.  Entropy is
highest where the surrogate mean prediction crosses $g = 0$, but it is high in
all areas where the 95\% interval captures $g = 0$ as well.  This is entropy's
desirable property -- it hones in on the failure contour.  Yet as we acquire
more points, entropy will become peakier and will
still favor acquisitions around the vertical purple bar.

Notice, there is a local optima in the entropy surface at the right-most
boundary where $x_1 = 1$. This arises primarily because the GP predictive
equations indicate higher uncertainty at the edges of the input space. Many AL
criteria are coupled with theory indicating that eventually such disconnected
``high uncertainty'' regions will recruit acquisitions because the criteria
will eventually be maximized there, after targeted sampling elsewhere saturates.
This is not true with entropy; it will never ``in-fill''.  A consequence of
this is that it will not explore the entire level set.  Once it finds a highly
probable crossing, like the one along the slice in the figure, it will only
explore near its current best estimate. In the parlance of AL/BO, it is too
exploitative, too myopic \citep{renganathan2021lookahead}.

There are several suggested solutions in the literature.  For example,
\citet{marques2018contour} integrate entropy over the input space
to target the largest reduction in global entropy rather
than simply the largest value.  This is a good idea in principle, but
unfortunately the integral is not available in closed form, and the quadrature
required is numerically fraught beyond 2d.  This approach does not compare
favorably to simpler, more {\em ad hoc} methods.  For example,
\citet{cole2022entropy} suggest deliberately coarsening the entropy search by
only finding local rather than global optima of the acquisition surface to
encourage exploration.  This works well, and represents the state-of-the-art
as far as we are aware.  However, it has the downside of requiring a tuning
parameter (trading off local versus global search) and derivative-based
optimization which is not amenable to MCMC-based surrogates.

\section{Non-stationary DGP Contour Location}\label{sec:method}
Here we describe our contribution: contour location (CL) with deep Gaussian
process (DGP) surrogates, triangulation candidates, and Pareto front
acquisitions.

\subsection{Deep Gaussian process surrogates}\label{sec:dgp}

DGPs are formed by layering stationary GPs. While this may be
accomplished with kernel convolutions \citep{dunlop2018how}, we prefer 
the lens of functional composition. A two-layer DGP prior is
defined as
\begin{equation}\label{eq:dgp}
\begin{aligned}
y_n \mid W &\sim \mathcal{N}_n\left(\mu_y, \Sigma(W)\right) \\
w_i &\stackrel{\mathrm{ind}}{\sim} \mathcal{N}_n\left(\mu_w, \Sigma(X_n)\right)
\quad \textrm{for}\;\; i = 1, \dots, p,
\end{aligned}
\quad\quad\textrm{where}\quad W = \begin{bmatrix} w_1 & w_2 & \dots & w_p \end{bmatrix},
\end{equation}
and $\Sigma(\cdot)$ is provided in Eq.~(\ref{eq:gpprior}) with potentially
unique hyperparameters.  Each $w_i$ is a conditionally 
independent stationary GP over the inputs
$X_n$.  Together these ``nodes'' form the latent $W$ layer, which serves as
a warped version of the inputs and feeds into an outer stationary GP
connecting to the response $y_n$.  The dimension of the latent layer need not
match the dimension of the input space, but $p = d$ has been shown to work
well \citep{sauer2023active}.  Prior means are adjustable, but 
common choices are $\mu_y = 0$ and either $\mu_w = 0$ or 
$\mu_w = X_n$ (the latter only being applicable when $p = d$).  We 
have found settings where both choices are advantageous, but here we 
opt for the simplest option $\mu_y = \mu_w = 0$.
Although additional GPs may be layered to create deeper models,
such complexity has been shown to provide marginal, if any, benefit for 
surrogate modeling tasks \citep{radaideh2020surrogate,sauer2023active} while 
requiring far more computation.  We thus focus on the two-layer case
(although we entertain a three-layer DGP in Figure \ref{fig:erf2results}).

Latent $W$ drives non-stationary flexibility.  Its nodes may ``stretch''
inputs in regions of high signal and ``squish'' inputs in regions of low
complexity. Yet the functional and multi-dimensional nature of this unknown
quantity poses an inferential challenge.  One solution is to approximate the
intractable DGP posterior with a ``close'' one (in terms of Kullback-Leibler
divergence) from a known target family; this is known as approximate
variational inference \citep[VI;][]{blei2017variational} and is popular in
machine learning applications where signal-to-noise ratios are low and data
are abundant \citep{damianou2013deep,salimbeni2017doubly}.  However, in our
deterministic, data-sparse surrogate modeling setting we opt for full posterior
integration to prioritize UQ.  We accomplish this using the Bayesian DGP
set-up of \citet{sauer2023active}. Latent $W$ are sampled through elliptical
slice sampling \citep[ESS;][]{murray2010elliptical}.  Kernel hyperparameters
are also sampled through MCMC, although these are more of a fine-tuning and
their sampling may be replaced with MLE-based alternatives
\citep[e.g.,][]{ming2022deep}. See \citet{sauer2023deep} for a thorough review
of Bayesian DGP surrogates.

\begin{figure}[ht!]
\centering
\includegraphics[width=17.5cm,trim=10 25 10 10]{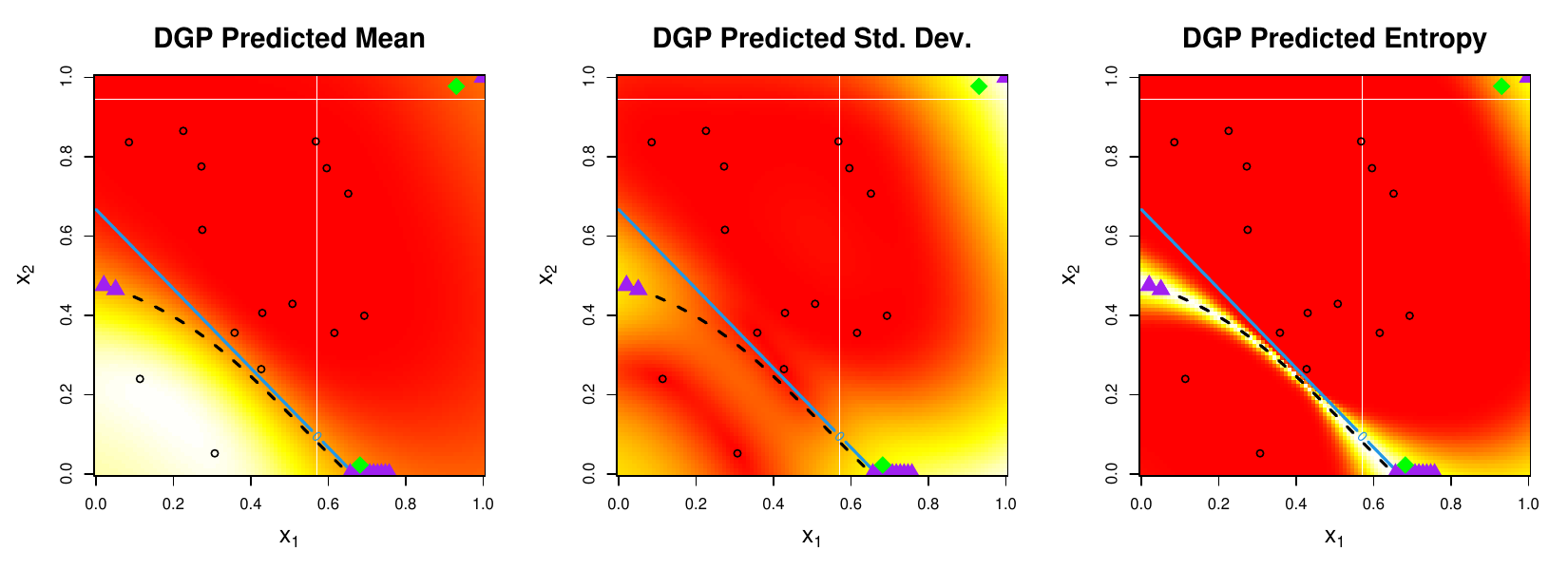}
\caption{Two-layer DGP mean (left), standard deviation (center), and 
{\it post hoc} entropy (right) for the plateau function (\ref{eq:plateau}): 
white/high, red/low.  Open circles indicate training data.  Purple triangles 
and green diamonds represent Pareto front acquisitions 
(Section \ref{sec:pareto}) under gridded candidates and 
``tricands'' (Section \ref{sec:tricands}), respectively.}
\label{fig:twoD}
\end{figure}

After burn-in, ESS provides posterior samples $w_i^{(t)}$ for $i = 1,\dots, p$
and $t\in\mathcal{T}$ (we shall drop reliance on kernel hyperparameters here,
which may also be indexed by $t$).  Samples in hand,
prediction at inputs $\mathcal{X}$ requires first ``warping'' the
predictive locations, $\mathcal{X}$ to $\mathcal{W}_i^{(t)}$, for each node
($i$) and each MCMC sample ($t$).  Each component layer is conditionally
Gaussian (\ref{eq:dgp}), so this means applying
Eq.~(\ref{eq:gppred}) with sampled $w_i^{(t)}$ in place of $y_n$.  We follow
\citet{sauer2023active} in mapping through the posterior predictive mean,
\begin{equation}\label{eq:dgpmap}
\mathcal{W}_i^{(t)} = \Sigma(\mathcal{X}, X_n)\Sigma(X_n)^{-1} w_i^{(t)}
\end{equation}
instead of drawing from the full Gaussian posterior.  Second, we column bind
$\mathcal{W}_i^{(t)}$ into the $n_p\times p$ matrix $\mathcal{W}^{(t)}$ and
again apply Eq.~(\ref{eq:gppred}) to obtain posterior moments
\begin{equation}\label{eq:dgpmoments}
\begin{aligned}
\mu_Y^{(t)} &= \Sigma(\mathcal{W}^{(t)}, W^{(t)})\Sigma(W^{(t)})^{-1} y_n \\
\Sigma_Y^{(t)} &= \Sigma(\mathcal{W}^{(t)}) - \Sigma(\mathcal{W}^{(t)}, W^{(t)})
\Sigma(W^{(t)})^{-1}\Sigma(W^{(t)}, \mathcal{W}^{(t)}).
\end{aligned}
\end{equation}
We thus obtain $|\mathcal{T}|$ samples for each posterior moment (typically
in the thousands).  It is often beneficial to summarize these quantities 
using the law of total variance,
\begin{equation}\label{eq:dgppred}
\mu_Y = \frac{1}{|\mathcal{T}|}\sum_{t\in\mathcal{T}} \mu_Y^{(t)}
\quad\textrm{with}\quad \Sigma_Y = \frac{1}{|\mathcal{T}|}\sum_{t\in\mathcal{T}}
\Sigma_Y^{(t)} + \mathbb{C}\mathrm{ov}(\mu_Y^{(t)})
\quad\textrm{and}\quad
\sigma_Y = \sqrt{\mathrm{diag}(\Sigma_Y)}.
\end{equation}
To provide a concrete visual, we fit a two-layer Bayesian DGP to the data from
Figure \ref{fig:oneD}. The right panel of that figure shows the DGP fit over
the 1d slice. Compared to the GP (middle panel), the DGP is able to more
accurately capture the plateau shape of the surface and the sloping drop
between the two flat regions.  The full 2d DGP mean surface is shown in the
left panel of Figure \ref{fig:twoD}, with training data locations indicated by
open circles.  The estimated failure contour is indicated by the black dashed
line; the true failure contour is overlayed in blue for reference.  The center
panel displays a heat map of the predicted standard deviation, $\sigma_Y$ from
Eq.~(\ref{eq:dgppred}). Notice that uncertainty is high in areas near the
transition and far from training data. The equivalent figure for the one-layer
GP is relegated to Supp.~C since posterior uncertainty for a
stationary GP is less interesting, being solely a function of distance to
training data.

Although the full DGP posterior is no longer strictly Gaussian, its components
are conditionally Gaussian, allowing us to utilize the failure probability calculation
of Eq.~(\ref{eq:p}) on component pieces to evaluate entropy for CL.  Using 
Eq.~(\ref{eq:dgpmap}--\ref{eq:dgpmoments}) with $\mathcal{X} = x$ and with MCMC-based 
expectation over $t$, we obtain
\begin{align}\label{eq:dgpent}
H(x) &= \frac{1}{|\mathcal{T}|}\sum_{t\in\mathcal{T}}
\left[ 
-p_x^{(t)}\log(p_x^{(t)})- (1\!-\!p_x^{(t)})\log(1\!-\!p_x^{(t)})
\right], & 
p_x^{(t)} &= 1 - \Phi\left(\frac{g - \mu_Y^{(t)}(x)}{\sigma_Y^{(t)}(x)}\right).
\end{align}
It may be more expedient to plug the summarized moments $\mu_Y$ and $\Sigma_Y$
from Eq.~(\ref{eq:dgppred}) directly into the Gaussian CDF of
Eq.~(\ref{eq:p}). We refer to this as {\it post hoc} entropy since it uses 
post-processed DGP moments. The right panel of Figure \ref{fig:oneD}
shows both entropy calculations along the 1d slice as red and dashed black
lines.  They are equivalent for the GP since there was no
warping at play.  
For the DGP, the shapes of the curves are very similar.  Absolute
magnitudes do not feature in our acquisitions; entropy values are only
considered relative to each other. Both acquisition surfaces identify
a single high-entropy region where the predicted surface crosses the failure
boundary. We have noticed this in all of our experiments (not shown here), and
so for simplicity and ease-of-use (the {\it post hoc} calculation is more
``plug-and-play''), we utilize {\it post hoc} entropy here on out.  However
our software supports both options.

The right panel of Figure \ref{fig:twoD} displays a heat map of the {\it post
hoc} entropy surface from the two-layer DGP fit to the plateau function. Notice
how it is very peaky, with a ridge of maxima along the entire predicted contour.
For now, ignore the purple triangles and green diamonds, which are discussed
later in Section \ref{sec:pareto}.  Suppose we sought a new
acquisition utilizing the entropy criterion, based on this DGP surrogate.
There are two clear obstacles.  First, finding optima in the entropy surface
comes with a hefty computational price-tag.  The ridge highlighted in Figure
\ref{fig:twoD} was found by evaluating the DGP surrogate on a dense grid, a
method that is infeasible in higher dimensions. 
Numerical optimizers, on the other hand, require many successive evaluations
of singleton predictive locations.  In a DGP, each predictive location must be
mapped through the latent layer (\ref{eq:dgpmap}), then to posterior
moments (\ref{eq:dgpmoments}), with this entire process repeated
$|\mathcal{T}|$-many times before final aggregate predictive mean and
variances are available (\ref{eq:dgppred}). Doing this evaluation
one-at-a-time (without closed-form gradients), is incredibly inefficient.  A
common workaround in AL \citep[e.g.,][]{gramacy2009adaptive} and BO
\citep[e.g.,][]{eriksson2019scalable} involves deploying a
limited, discrete set of candidates.  With candidates, the loop over
$t\in\mathcal{T}$ need only be done once.  Although simple to implement, the
fidelity of such candidate searches is low in situations where the goal is to
identify a small target (e.g., a contour) in a big input space.  Space-filling
candidates, like LHS's, are likely to miss peaky areas of high entropy, unless
the size of the candidate set is prohibitively large.

The second obstacle involves the entropy criterion itself; it is too myopic
for CL \citep[see, e.g.,][]{marques2018contour,cole2022entropy}. Even if we
are able to find optima in the entropy surface, this criterion only minimally
accounts for predictive uncertainty.  In Figure \ref{fig:twoD}, a
portion of the high entropy ridge is actually very close to existing training
data locations.  Even though uncertainty is low here, entropy is still high.
Acquisitions near the outer edge of the predicted contour where uncertainty
is higher, or in the upper-right corner where there is no training data,
would be better.  Although there is a local maximum in the entropy surface in
the upper-right corner, this local maximum will never overcome the global
maximum near the predicted contour.  This is why the hybrid-local-optimization
scheme of \citet{cole2022entropy}, which successfully identifies local optima
in the entropy surface, outperforms globally-optimized competitors.  But
further mitigation of entropy's limitations is warranted.

\subsection{Triangulation candidates}\label{sec:tricands}

\begin{figure}[ht!]
\centering
\includegraphics[width=17.5cm,trim=10 25 10 10]{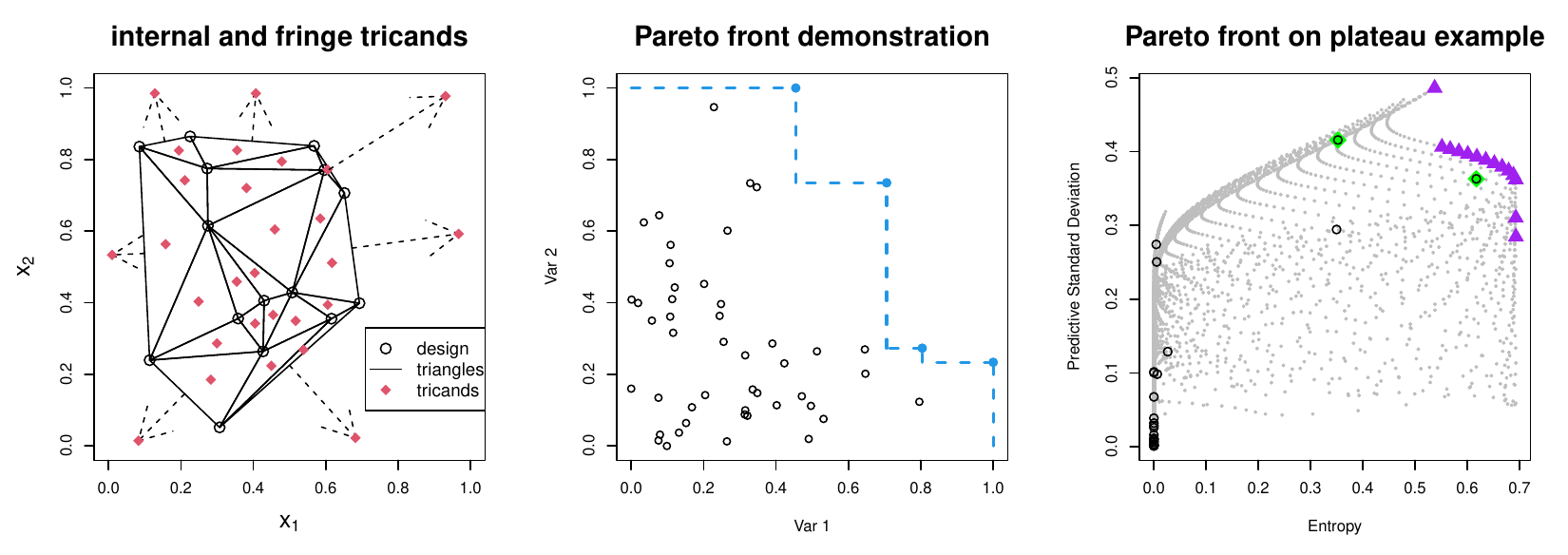}
\caption{{\it Left:} Delaunay triangulation with internal and fringe ($\alpha
= 0.9$) ``tricands''. {\it Middle:} A Pareto front 
set (blue dashed lines) for two hypothetical variables.  {\it Right:} Entropy
versus standard deviation for the DGP of Figure \ref{fig:twoD} two ways: on a
grid in $x$-space (gray dots), with the Pareto front as purple triangles; via
``tricands'' (black open circles) with Pareto front as green diamonds.}
\label{fig:Pareto}
\end{figure}

Here we address the first issue, and to a lesser extent the second (which is
more squarely the subject of Section \ref{sec:pareto}), by adapting a new
candidate scheme from the BO literature: triangulation candidates
\citep{gramacy2022triangulation}, or ``tricands''.  Rather than allocating
candidates at-random or in a space-filling scheme, we allocate candidates
strategically by ``in-filling'' the gaps in the existing design $X_n$. Start
with a {\it Delaunay triangulation} of $X_n$: a set of line segments between
points that divy up the space so no lines cross.  The left panel of Figure
\ref{fig:Pareto} provides a visual of this in 2d (for the training data
from the 2d plateau function example). The other panels will be narrated
in Section \ref{sec:pareto}.   Open circles indicate observed training data
locations, and black solid lines form the Delaunay triangulation.  In two
dimensions, this triangulation forms triangles; in higher dimension it forms
tetrahedra although we will still refer to them as triangles.  Each triangle
is defined by a $(d + 1) \times d$ matrix.  Denote each triangle by $T_j$ for
$j = 1, \dots, n_T$.  The number of triangles ($n_T$) is affected by the size
of the training data ($n$) and the dimension of the input space ($d$).

Tricands are allocated based on the triangulation in two realms: internal and
fringe.  Internal candidates are located at the {\it barycenter} of each
triangle: $\tilde{x}_j = \frac{1}{d+1}\sum_{i=1}^{d+1} T_j[i, \;]$.
Notice in Figure \ref{fig:Pareto} how they spread out between existing design points.
For fringe candidates, denote $F_j$ for $j = 1, \dots, n_F$ as the $d\times d$
matrix representing a facet (outer edge of the triangulation). 
Calculate the midpoint of each facet as
$\bar{F}_j =
\frac{1}{d}\sum_{i=1}^d F_j[i, \;]$ for $j = 1,\dots,n_F$. Then take the
normal vector $\vec{v}_j$ extending perpendicularly from $F_j$ at the midpoint
$\bar{F}_j$ (dashed arrows in Figure \ref{fig:Pareto}, left). Allocate a
fringe candidate along each $\vec{v}_j$, with the exact placement determined
by a tuning parameter $\alpha\in[0, 1]$ representing the proportional distance
from the facet midpoint to the boundary of the space. For BO,
\citet{gramacy2022triangulation} found $\alpha = 0.5$ worked well. For CL,
we choose $\alpha = 0.9$ to encourage acquisitions near the boundary --
contours are often found near boundaries, especially in reliability
applications. Fringe candidates  are represented by the tips of dashed arrows
in the figure. Crucially, tricands (both internal and fringe) are agnostic to
surrogate models; they rely only on the geometry depicted by existing training
locations, spreading out to encourage exploration while still offering a
higher concentration of candidates in more densely observed areas to allow for
exploitation.

Denote the combined set of internal and fringe tricands as $\mathcal{X}_N$,
where $N = n_T + n_F$.  As input dimension ($d$) and data size ($n$) increase, 
the total number of tricands will grow.  When $N$ is large, it
is helpful to limit the computational burden by setting a maximum size of the
candidate set, $N_\textrm{max} \ll N$.  While a rudimentary approach would
simply sub-sample $N_\textrm{max}$ points from all $N$ tricands, we find it
helpful to employ a targeted sub-sampling approach to guarantee retention of
some exploitave candidates near the failure contour.  This approach is
inspired by the BO-based sub-sampling of \citet{gramacy2022triangulation}, but
it requires modification to suit the CL setting.  Rather than retaining
candidates around a {\em single} best-observed point as in BO,
we seek to retain 
candidates around the entire contour.  Since the true contour is
unknown, we leverage the absolute difference between the observed responses
and the limit state ($|y_i - g|$) to rank the observations by closeness to
the contour.
Then, starting with the observation closest to $g$, we identify all triangles $T_j$
and/or facets $F_j$ that were generated with this point, and randomly select
one of their candidates to retain.  We repeat this process, working down the
ordering and retaining one candidate {\it adjacent} to each location, until we
have collected at least 10\% of $N_\textrm{max}$.  Then we select remaining
candidates, filling out $N_\textrm{max}$, completely at random.  By
iteratively working down the ordering of closeness to the contour, we are able
to target the entire contour set.  As with ordinary tricands, this targeted
sub-sampling does not rely on any surrogate information.

\subsection{Pareto front acquisitions}\label{sec:pareto}

Tricands circumvent a continuous, derivative-based optimization (which is troublesome
with MCMC-based surrogates), by replacing it with a geometric calculation.
They also, to a degree, encourage exploration of the acquisition space by
placing candidates where predictive uncertainty is likely to be high. However,
we have found that this is not enough to correct the myopia of entropy.
Toward that end, we propose an acquisition criterion that strikes a deliberate
balance between entropy (exploitation) and predictive uncertainty
(exploration).  Although approaches similar to the ones we describe have been
suggested in other AL contexts, we are not aware of any such attempts for 
CL, or with tricands or similar candidate-based schemes.

Consider entropy $H(x)$ and predictive standard deviation $\sigma_Y(x)$
evaluated over $x \in \mathcal{X}$ in the input space.  When using either
as a lone acquisition criterion, higher is better.  Now let $c(x) = (H(x),
\sigma_Y(x)) \in \mathbb{R}^2$ represent a two-dimensional ``criterion'' and write
$c(x) \prec c(x')$, for $x' \in \mathcal{X}$, if $c(x')$ {\em strictly
dominates} $c(x)$, meaning that it is better in both dimensions
simultaneously: $H(x) < H(x')$ and $\sigma_Y(x) < \sigma_Y(x')$.  The {\em
Pareto frontier}, 
defined as $P(\mathcal{X}) = \{ x \in \mathcal{X} : \{ x'
\in \mathcal{X} : c(x) \prec c(x'), \, x \ne x' \} = \emptyset \}$, is the
subset of the input space $\mathcal{X}$ separating dominated from non-dominated
points.  The definition is similar for higher-dimensional criteria; for more details see \citet[][Chapter
4]{goodarzi2014introduction}.\footnote{Note that most textbook definitions of
Pareto fronts/non-dominated sets do not involve an input, $x$-variable,
working instead on generic $c$-criterion values.  However, we have introduced
$x$ to make an explicit link to criteria evaluated over an input space
$\mathcal{X}$.  To connect to classical
definitions one could instead write $c(P(\mathcal{X}))$.}

The Pareto frontier is a compact set, and in surrogate modeling/AL contexts it
is possible to estimate and quantify uncertainty around it
\citep[e.g.,][]{binois2015quantifying,luo2018evolutionary} for arbitrary $x
\in \mathcal{X}$.  When restricting to a discrete subset of the input space,
like to tricands $\mathcal{X}_N$, the Pareto front $P(\mathcal{X}_N)$ is also
discrete, and is known as the {\em non-dominated set}.  
For example, consider random samples on $c(x) \in \mathbb{R}^2$, displayed as
{\tt Var1} and {\tt Var2} in the middle panel of Figure \ref{fig:Pareto}.  We
generated these directly in $c-$space, without generating $x$'s first, for a
simple illustration, but we shall return to $x$-space momentarily.  The blue 
filled circles connected by the dashed staircase indicate the points on the
Pareto front, $P$ (assuming larger is preferred).  
These are not beaten by any other point with respect to both
variables: other points may be higher on one or the other, but not both.

Identifying the Pareto front from a discrete sample in 2d is
straightforward: (a) order the observations from highest to lowest for one
variable (points that yield the maximum of either variable are always in the
Pareto front set); and (b) work down the ordering one-by-one, adding points to
$P$ only if they are higher on the second variable than all
preceding points (which were higher on the first variable).  Our
implementation is 5--6 lines of {\sf R} code, available in our Git repository.
However we note that the {\tt rPref} {\sf R}-package on CRAN \citep{rPref}
offers faster computations for larger datasets in higher dimensions.  

For a particular example in our CL setting, let us return to the plateau
surface of Figure \ref{fig:twoD}, which involves two input dimensions, and now
two acquisition criteria $c(x) = (H(x), \sigma_Y(x))$.  The right panel of
Figure \ref{fig:Pareto} combines the visuals from the center and right panels
of Figure \ref{fig:twoD} by plotting $H$ versus $\sigma_Y$, evaluated over a
dense grid of locations (gray dots), and also over tricands $\mathcal{X}_N$ 
(open black circles).  An ideal acquisition would be in the upper right corner of
this plot, with both high entropy and high uncertainty.  In both Figures
\ref{fig:twoD} and \ref{fig:Pareto}, candidates along the Pareto front are
highlighted by purple triangles for the candidate grid, and green diamonds for
tricands.  Figure \ref{fig:twoD} is providing the $x$-location of the Pareto
front(s), whereas Figure \ref{fig:Pareto} shows the corresponding
$c(x)$-value(s).

Observe in the Figure(s) that tricands provide similar information at a
fraction of the cost compared to a dense grid. 
Any of the points on either Pareto front (green diamonds or
purple triangles) represents a sensible acquisition.   
Observe in Figure \ref{fig:twoD} that all of
these choices are far from the training data $X_n$, thus correcting the mypoia
of entropy.  They are either very close to the estimated contour, but away
from $X_n$, or if they are not near the contour they are {\em very} far
from $X_n$. We prefer to select an acquisition completely
at random from $P(\mathcal{X}_N)$. Going forward we use only tricands
because grids that are dense enough for good resolution become
prohibitively large in higher input dimension.

It is worth noting that tricands have an extra advantage over grids -- 
in $x$-space, the members of $P(\mathcal{X}_N)$ (green diamonds) are located far 
from the training design $X_N$ {\em and} far from one
another by virtue of their geometric nature.  This is not true for Pareto fronts
constructed from a dense grid.  This means a tricands--Pareto set could be
used to select a batch of new runs at once \citep{cole2022entropy}. We
provide further discussion of this idea in Section \ref{sec:discuss}.

To summarize, our candidate-based contour location scheme boils down to three
steps: (1) build tricands $\mathcal{X}_N$ from $X_n$; (2) fit a DGP surrogate to
evaluate posterior predictive standard deviation $\sigma_Y(\mathcal{X}_N)$ and
predictive entropy $H(\mathcal{X}_N)$ for these candidates; (3) select an
acquisition uniformly at random on the Pareto front of uncertainty and
entropy: $x_{n+1}
\sim \mathrm{Unif}[P(\mathcal{X}_N)$]. Supp.~D provides 
a summary of this algorithm.  While this 
Pareto--tricands CL scheme is
technically independent of surrogate modeling choices and may be deployed with
any surrogate -- and also any multi-dimensional acquisition criteria -- we
find DGPs are crucial for the non-stationary response surfaces that motivate
this work. Empirical evidence  is presented in Sections
\ref{sec:results}--\ref{sec:su2}.

\section{Implementation and Benchmarking}\label{sec:results}

In this section we describe the publicly available implementation
of our method, then validate it on a variety of synthetic test problems.

\subsection{Implementation details}\label{sec:implement}

Code to reproduce all results (Sections \ref{sec:synthetic}--\ref{sec:su2}) 
is provided in our public Git 
repository.\footnote{\url{https://bitbucket.org/gramacylab/deepgp-ex/}}
We utilize the {\tt deepgp} package for {\sf R} on CRAN \citep{deepgp} for
Bayesian DGP surrogate training and prediction.  That package was recently
updated to provide entropy calculations following Eq.~(\ref{eq:dgpent}).
Even though full entropy calculation is offered in the package, we opt for the {\it post hoc}
calculations using summarized posterior moments.  All experiments
utilize package defaults aside from fixing the
noise parameter $\eta = 1\times 10^{-6}$ 
for interpolation of deterministic computer models.  

We use the {\tt geometry} \citep{geometry} {\sf R}-package for
Delaunay triangulation, which is based on a legacy {\sf C} library
called {\tt quickhull} \citep{quickhull}.  We offer convenient {\sf R} 
and {\sf python} wrappers which include calculation of internal and fringe
``tricands'' with the targeted sub-sampling approach
described in Section \ref{sec:tricands}.
Identification of the Pareto front requires a simple {\tt for} loop 
over the ordered observations and an {\tt if} statement to check values
of the second variable.  
Our main performance metric is sensitivity
(Supp.~B) on a hold-out LHS testing set over iterations of
acquisition, or for a single space-filling LHS of fixed size.  Higher
sensitivity is better, with  1.0 indicating detection of all failures.
Discussion of computation times is reserved for Supp.~E.

We entertain the following surrogates as competitors:
\begin{itemize}
	\item DGP ESS: two-layer Bayesian ESS-based DGP of \citet{sauer2023active} using
	{\tt deepgp} in {\sf R}	
	\item GP MCMC: stationary GP with MCMC-sampled 
	separable lengthscales, also via {\tt deepgp}
	\item GP MLE: stationary GP with maximum likelihood
	estimated separable lengthscales using {\tt scikit-learn} in {\sf Python}
	\citep{scikit}
	\item DGP VI: doubly stochastic VI DGP of
	\citet{salimbeni2017doubly} using {\tt gpytorch} in {\sf Python}
	\citep{gpytorch}
\end{itemize}
All utilize the Mat\`ern $\nu = 5/2$ kernel, and all training data
are observed without noise.   
For each model, we compare a static space-filling
LHS to a strategic CL sequential design (when feasible). We deploy our
tricands-Pareto-front acquisition scheme with the MCMC-based surrogates (DGP
ESS pareto and GP MCMC pareto).  Since these two differ only in the choice of
surrogate (two-layer DGP versus stationary GP), they offer a glimpse into the
direct impact of non-stationary flexibility on CL performance. We deploy the
hybrid entropy-based numerical optimization scheme of \citet{cole2022entropy}
with the MLE-based GP (GP MLE hyb ent), which is the state-of-the-art at the
time of publication, and thus our main competitor.

The DGP VI surrogate employs approximate VI through a stochastic optimization
of the evidence lower bound. This
optimization entails several tuning parameters including the learning
rate, number of optimization steps (epochs), and the number of Gaussian
mixtures in the target posterior.  There exist automated tuning methods 
involving inner-optimizations \citep[e.g.,][]{zimmer2021auto}; however, 
these approaches incur a significant computational expense. In the context of 
sequential design for CL, having an internal sequential design (i.e., a BO) 
to fine-tune optimization parameters seems to us like a rabbit hole.
Rather, we fix these values at what we consider to be good defaults 
(mimicking the other surrogates which all use package defaults):
$1\times 10^{-3}$ learning rate, 2500 epochs, and 32 Gaussian mixtures.
We include the DGP VI competitor as an alternative DGP benchmark,
but admit this comparison is tangential to our main objective.  In light of 
this, and DGP VI's poor performance, we only entertained it in the static settings.

\subsection{Synthetic examples}\label{sec:synthetic}

\subsubsection{Plateau Function}  
\begin{figure}[ht!]
\centering
\includegraphics[width=16cm,trim=10 27 10 10]{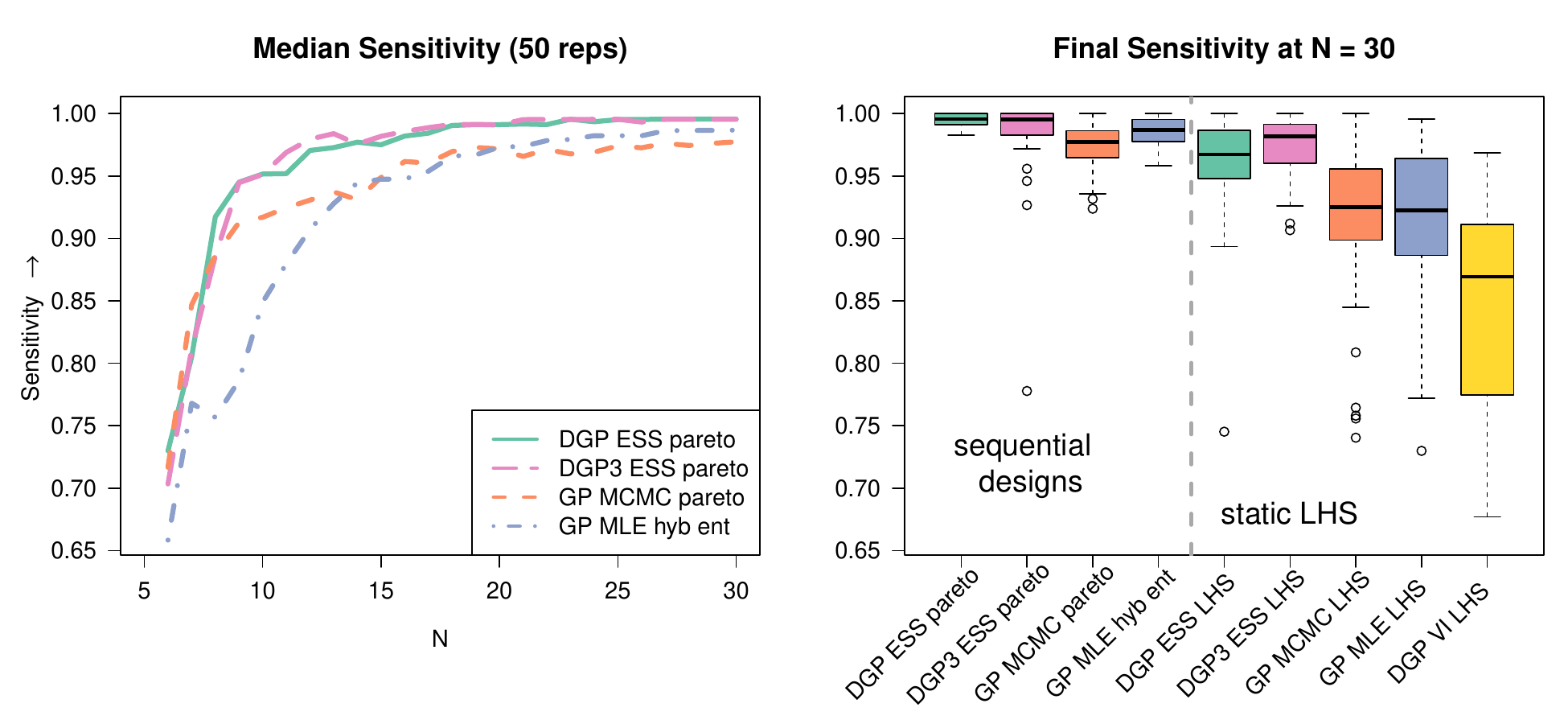}
\caption{Sensitivity (higher is better) for the 2d plateau function.}
\label{fig:erf2results}
\end{figure}

Consider the following function,
\begin{equation}\label{eq:plateau}
f(x) = 2 \cdot \Phi\left(\sqrt{2} \left(-4 - 3\sum_{i=1}^d x_i\right)\right) - 1
\quad\textrm{for}\; x\in[-2, 2]^d \; \left(\textrm{scaled to}\; x\in[0, 1]^d
\right),
\end{equation}
which we have adapted from \citet{izzaturrahman2022modeling} to be defined in
arbitrary dimension.  This function formed the basis of our earlier
illustrations in Figures \ref{fig:oneD}--\ref{fig:Pareto}. We define the contour
at $g = 0$ such that $f(x) > 0$ indicates a failure. Starting with $d = 2$
and an initial LHS design of size $n_0=5$, we conduct sequential designs
targeting this contour up to an ending design size of $n = 30$ (25
acquisitions).  Median sensitivity across 50 Monte Carlo (MC) repetitions with
re-randomized starting designs is displayed in the left panel of Figure
\ref{fig:erf2results}.  Sensitivities after the last acquisition are 
displayed in the left interior panel.  
As an additional benchmark, we fit each surrogate on
static LHS designs of equivalent size ($n = 30$) and reported the
sensitivities in the right interior panel.  For this example, we also 
entertained a three-layer DGP (DGP3) using {\tt deepgp} with package defaults.

Although the DGP surrogate starts at a similar place as the MCMC-based GP, it
quickly jumps into the top position and maintains this superiority for the extent of the design.  
Median performance of the three-layer DGP matched that of the two-layer,
but the additional flexibility of the deeper model led to poorer worst-case
performance (which we attribute to over-fitting).  The three-layer DGP required twice
the compute time of the two-layer (Supp.~E); prompting us to
drop it from further consideration. All methods
benefited from CL acquisition when compared to their static LHS counterparts.
The DGP VI model struggled to cope with this small-data setting; it was
designed for much larger problems.

\begin{figure}[ht!]
\centering
\includegraphics[width=16cm,trim=10 27 10 10]{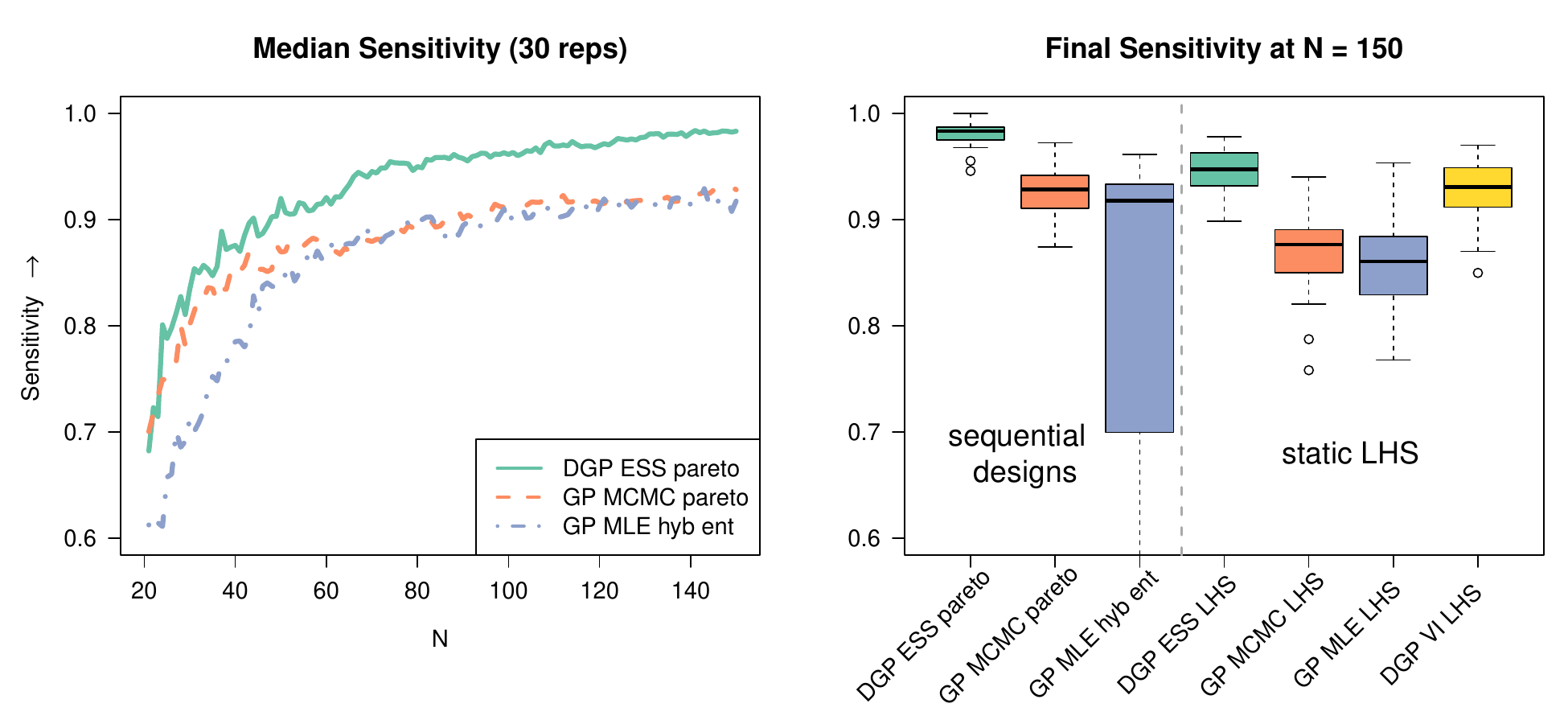}
\caption{Sensitivity for the 5d plateau function.}
\label{fig:erf5results}
\end{figure}

Next, we bumped the dimension up to $d = 5$ and
repeated this exercise, starting with $n_0 = 20$ and acquiring up to $n = 150$.  
Results are shown in Figure \ref{fig:erf5results}. Performance trends are similar to those
of Figure \ref{fig:erf2results}, although it appears the DGP has an even
bigger edge in this higher-dimensional setting.  Also noteworthy,
entropy-based acquisitions behind ``GP MLE hyb ent'' occasionally led the
GP astray and resulted in poorer performance than their static design
counterparts, a byproduct of clustered acquisitions.   
DGP VI LHS performs better, comparatively, than in the 2d setting.  
We attribute this to the larger training data size.

\subsubsection{Cross-in-tray function}  
\begin{figure}[ht!]
\centering
\includegraphics[width=16.5cm,trim=10 25 10 10]{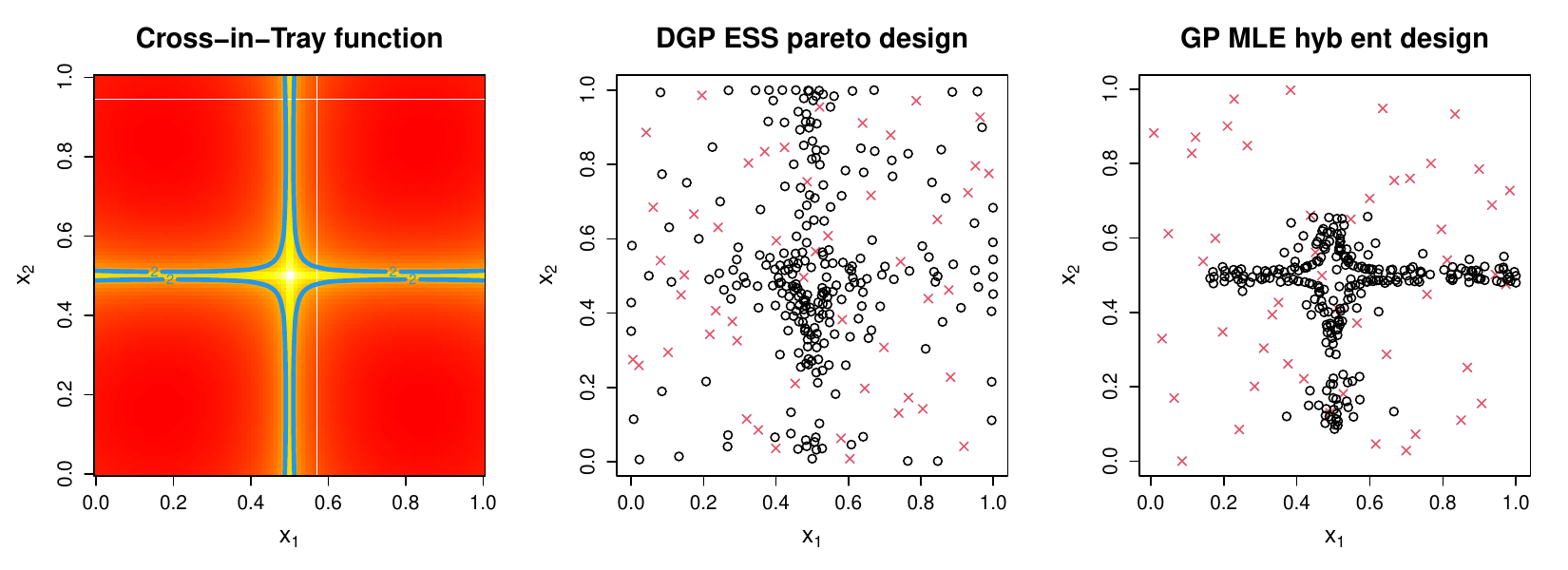}
\caption{{\it Left:} Heat map of 2d cross-in-tray function with 
contour at $g = 2$.
{\it Center:} Sequential design from two-layer DGP with Pareto-tricands acquisitions.  
{\it Right:} Sequential design from stationary GP with 
hybrid-entropy acquisitions.  Red ``x''
indicate starting LHS; black circles indicate acquired points.}
\label{fig:tray2}
\end{figure}

\begin{figure}[ht!]
\centering
\includegraphics[width=16cm,trim=10 27 10 10]{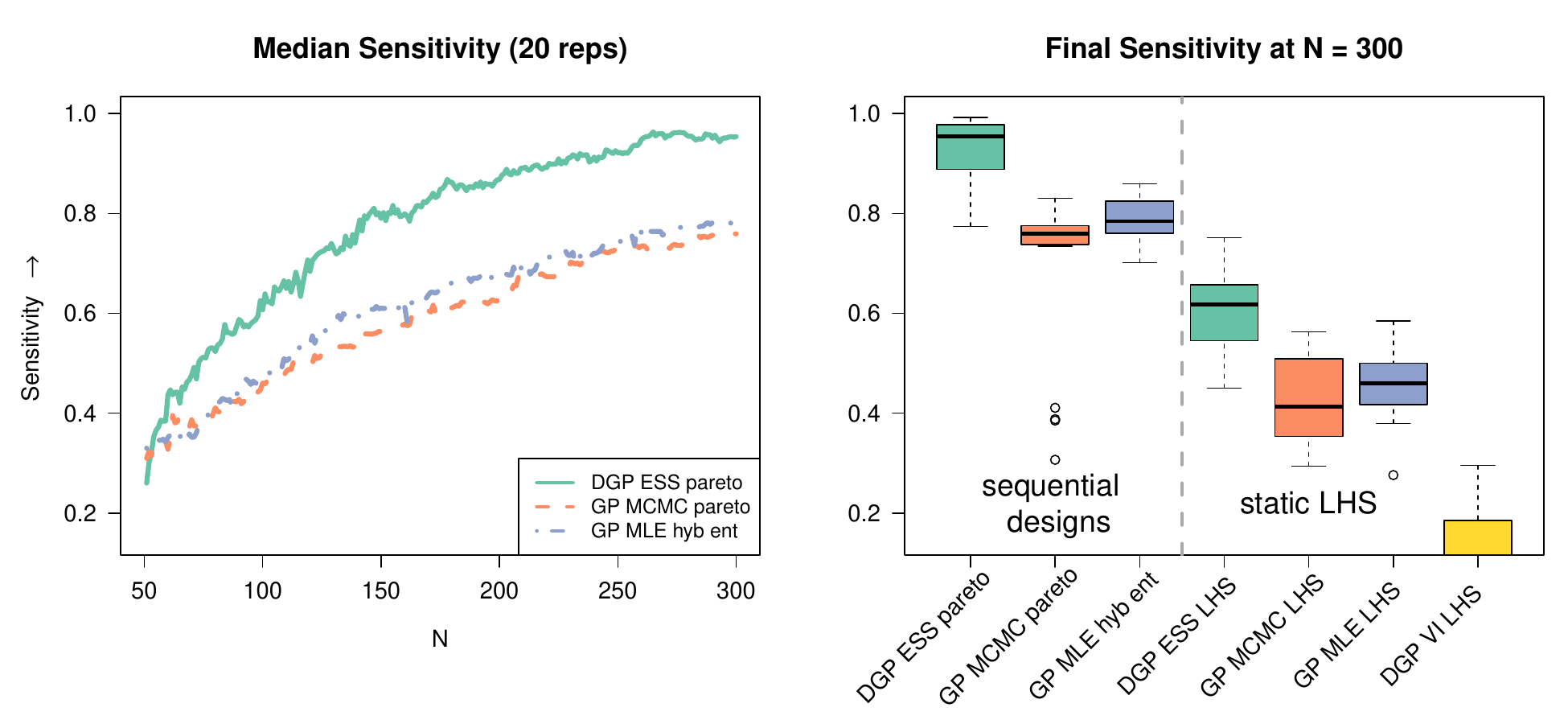}
\caption{Sensitivity for the 2d cross-in-tray function.}
\label{fig:tray2results}
\end{figure}

Next, consider the ``cross-in-tray'' function,
\[
f(x_1, x_2) = -0.001\left(\left|\sin(x_1)\sin(x_2)\mathrm{exp}
	\left(\left|100 - \frac{\sqrt{x_1^2 + x_2^2}}{\pi}\right|\right)\right| + 1 \right)^{0.1}
\]
found on the pages of the Virtual Library of Simulation Experiments
\citep{surjanovic2013virtual}.  We use the
domain $x\in[-2, 2]^2$ (similarly to $[0, 1]^2$).  A visual of the
surface is provided in Figure \ref{fig:tray2} (left) with a contour
defined at $g = 2$.  
We follow the same sequential design procedures, but we
increase training data sizes to $n_0 = 50$ with a final design size of $n =
300$ to account for the increased complexity in the surface and the contour.
Results over 20 MC repetitions are displayed in Figure \ref{fig:tray2results}.
Again, all methods benefit from sequential design, but the DGP has a clear
advantage due to its non-stationary flexibility.  The poor performance of DGP
VI LHS here is intriguing; perhaps it could be improved with fine-tuning. But
we suspect that the built-in inducing point approximations
\citep{snelson2006sparse}, which have been shown to hinder approximation
fidelity \citep{sauer2022vecchia}, are restricting the model's ability to
handle the finer scale of this failure region.

In addition, the center and right panels of Figure \ref{fig:tray2} display the
design at the end of a single MC exercise for the ``DGP ESS pareto'' 
and the ``GP MLE hyb ent'' schemes.  Red ``x'' indicate the starting LHS
design, and black circles indicate acquired points.  As expected,
acquisitions that rely solely on entropy tend to cluster (right panel).
Pareto front acquisitions do a better job of exploring the contour.  The real
``secret sauce'' though is the combination of the Pareto-tricands acquisition
scheme with the flexible DGP model.

\section{RAE-2822 Airfoil Contour Location}\label{sec:su2}

In Section \ref{sec:motive}, we previewed surrogate predictive
prowess on space-filling LHS designs of size $n=500$.  Results were shown 
earlier in Figure \ref{fig:su2static}, where we observed that the Bayesian 
DGP indeed outperformed the stationary GP. The performance of DGP VI was 
underwhelming; again we suspect fine-tuning may help, 
but the main culprit is a blurry inducing point approximation.

\begin{figure}[ht!]
\centering
\includegraphics[width=18cm,trim=10 20 10 10]{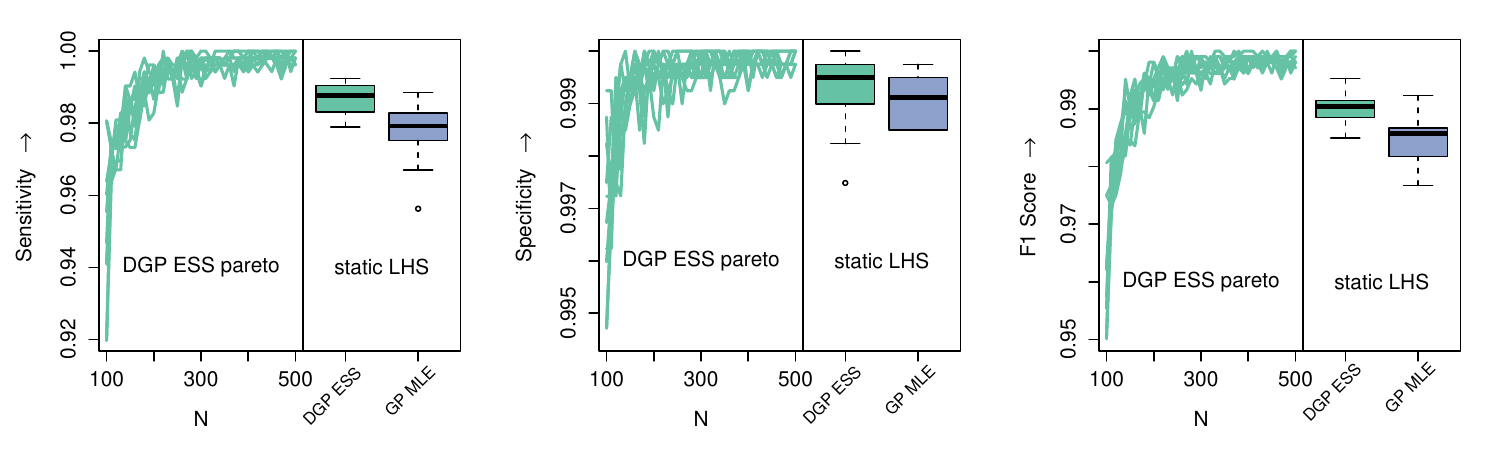}
\caption{Sensitivity, specificity, and F1 score 
for ``DGP ESS pareto'' 
sequential designs of the 7d airfoil simulation (left interior panels).  
Right interior panels show results from static LHS designs (DGP VI omitted).}
\label{fig:su2sequential}
\end{figure}

To improve performance with fewer simulations, we deployed our Pareto-tricands
AL scheme for CL with the DGP ESS surrogate. We initialized with a $100$-point LHS
and acquired 400 points for a resulting design of size $n
= 500$.  Sensitivity, specificity, and F1 scores (Supp.~B, higher is
better) for 10 MC repetitions (with re-randomized initial and testing LHS
sets) are shown in Figure \ref{fig:su2sequential}.  The left interior panels
show progress across the sequential designs, and the right interior panels
show the results from static designs of equivalent size (copied from Figure
\ref{fig:su2static} for reference). DGP-tricands-Pareto sequential designs
outperformed static fits with as few as 200 evaluations of the simulator, a
savings of nearly 12 hours of compute time compared to the 500-point LHS.

\begin{figure}[ht!]
\centering
\includegraphics[width=9.5cm,trim=10 10 10 10]{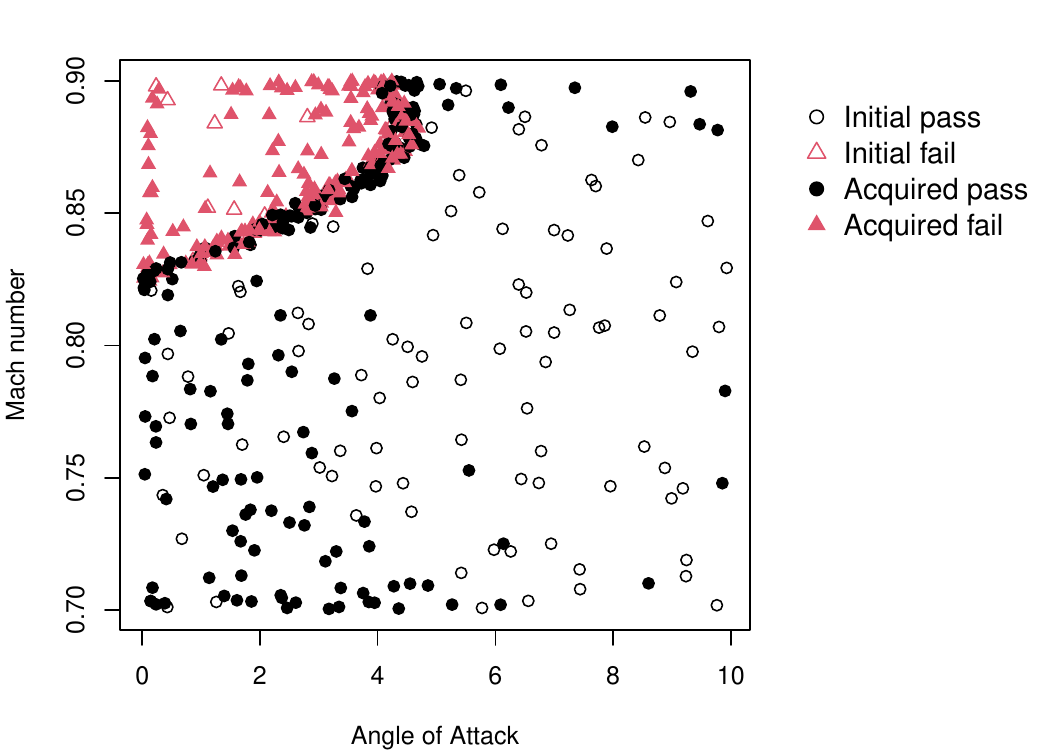}
\caption{2d projection of a ``DGP ESS pareto'' sequential design
($n = 500$).  Failures at $L/D < 3$.}
\label{fig:su2design}
\end{figure}

To help explain/visualize how the DGP ESS Pareto-tricands scheme is gaining
an advantage, Figure \ref{fig:su2design} shows one of the resulting sequential
designs, visualized as a projection over the two most impactful inputs: angle
of attack and Mach number.  
Pareto-tricands acquisitions (filled circles and
triangles) occasionally explore, but overwhelmingly focus on low angles of
attack.  All failures occurred at low angles of attack and high speeds.  This
makes sense, as this is where lift is generally low. Acquisitions
were often placed near this boundary in the upper left corner, an indication
that the AL procedure honed in on the contour.

\section{Discussion}\label{sec:discuss}

We proposed a sequential design scheme for contour location with DGPs 
that relies on triangulation candidates and selects acquisitions on
the Pareto front of entropy and posterior predictive uncertainty.  Our 
scheme circumvents cumbersome numerical optimizations, avoids
clustered acquisitions, and allows for CL with Bayesian DGPs (which has not
been done before).  While we were motivated by the application of
DGPs in aeronautic simulations, our sequential design scheme is not limited to
these cases.  Pareto-tricands sequential designs may be useful for other hefty
surrogate models which are incompatible with numerical optimization of
acquisition functions, such as an MCMC-based treed GP
\citep{gramacy2008bayesian}.  We believe that the Pareto front acquisition
procedure is applicable to any combination of multiple criteria.  One
advantage over other hybrid schemes, such as those based on aggregation, is
that it is independent of the relative scales of the component criteria.
Combining any criteria with predictive uncertainty and selecting acquisitions
on the Pareto front has potential to encourage exploration while retaining
exploitative behavior.

Our DGP-Pareto-tricands approach excels in settings where the failure
contour itself exhibits complex or non-stationary behavior.  Our synthetic
examples were designed with this aim in mind: the contour of the plateau function
is near the steep drop between regimes, and the contour of the cross-in-tray function
surrounds the non-stationary peaks of the surface.  If the response surface
is stationary, then the complexity of a DGP is not warranted and could even lead
to over-fitting.  GP diagnostics
\citep{bastos2009diagnostics} may help identify these situations.  Furthermore,
if the failure region is very small and/or the contour itself is rather simple,
a stationary GP may suffice, even if there is non-stationarity away from the
contour.  In these scenarios, the Pareto-tricands acquisitions designed to promote 
exploration may underperform simpler alternatives, like the hybrid entropy 
optimization of \cite{cole2022entropy}.

While tricands proved useful in our modest dimensional cases, the
computational costs of the Delaunay triangulation may become prohibitive in
higher dimensions.  In our experience, 8 dimensions is pushing the boundary of
tractability for the underlying {\tt quickhull} library, unless the design
size $n$ is kept low.  There are some opportunities for improvement, such as
only updating the portion of the triangulation near a newly acquired point.
But we suspect that new methodology will be needed to replace the Delaunay
triangulation in higher dimension while still mimicking its behavior.  One 
advantage to tricands/Pareto acquisition, which was not explored in this paper,
involves batch acquisition.  We observed in Section \ref{sec:pareto} that
it would be easy to take a batch whose size matched $|P(\mathcal{X})|$, the
size of the non-dominated set.  We note that one could always get more, to
augment the batch, by removing those points and recalculating the Pareto
front.  Exploring the extent to which this is a productive use of resources
would be an interesting subject of future research.

Reliability analysis involves quantifying the probability of failure given
uncertain inputs \citep[e.g.,][]{allen2004reliability}.  Our goal has been to
train a surrogate to accurately identify an entire failure contour.  The
surrogate can then be utilized with any probability distribution over the
input variables to estimate probabilities of failure.  Nevertheless, it may be
advantageous to incorporate information about a given input distribution into 
the sequential design procedure, so the surrogate can prioritize learning 
around certain regions of the contour.  \citet{abdelmalek2022bayesian} have shown
such methods to be advantageous with GPs, albeit at great computational expense.  We 
suspect similar strategies will work with DGPs, as long as the computational
burden can be kept in check.

\section*{Acknowledgements}
SAR acknowledges the allocation on Pennsylvania State University’s Institute 
for Computational and Data Sciences’ Roar supercomputer to run the simulator experiments.  RBG acknowledges partial support from NSF 2318861 and 2152679.

\bibliography{main}
\bibliographystyle{jasa}

\newpage
\appendix
\begin{center}
{\Large\bf SUPPLEMENTARY MATERIAL}
\end{center}

\section{Acronyms}

The following table offers a complete list of acronyms used throughout this manuscript.

\medskip

\begin{tabular}{l l}
AL & active learning \\
BO & Bayesian optimization \\
CFD & computational fluid dynamics \\
CL & contour location \\
CRPS & continuous ranked probability score \\
DGP & deep Gaussian process \\
ESS & elliptical slice sampling \\
GP & Gaussian process \\
L/D & lift / drag \\
LHS & Latin hypercube samples \\
MCMC & Markov chain Monte Carlo \\
MLE & maximum likelihood estimation \\
SUR & stepwise uncertainty reduction \\
RMSE & root mean squared error \\
UQ & uncertainty quantification \\ 
VI & variational inference \\
\end{tabular}

\section{Evaluation metrics}

\paragraph{Pass/Fail Classification Performance.} For a 
given observation $f(x)$ with failures defined at $f(x) > g$
and surrogate posterior predicted
mean $\mu_Y(x)$, denote the classification of $x$ as
\[
C(x) = \begin{cases}
    \textrm{TP (true positive)}, & f(x) > g \;\;\textrm{and}\;\; \mu_Y(x) > g \\
    \textrm{FP (false positive)}, & f(x) \leq g \;\;\textrm{and}\;\; \mu_Y(x) > g \\
    \textrm{TN (true negative)}, & f(x) \leq g \;\;\textrm{and}\;\; \mu_Y(x) \leq g \\
    \textrm{FN (false negative)}, & f(x) > g \;\;\textrm{and}\;\; \mu_Y(x) \leq g \\
\end{cases}
\]
Alternatively, swap all $>$ and $<$ signs if failures are instead
defined at $f(x) < g$.
For an entire set of locations $\mathcal{X}$, let $|\textrm{TP}|$ represent the
size of the set of rows of $\mathcal{X}$ classified as TP (and so on).  Then
\[
\mathrm{Sensitivity} = \frac{|\textrm{TP}|}{|\textrm{TP}| + |\textrm{FN}|},
\quad
\mathrm{Specificity} = \frac{|\textrm{TN}|}{|\textrm{TN}| + |\textrm{FP}|},
\quad\textrm{and}\quad
\textrm{F1 score} = \frac{2|\textrm{TP}|}{2|\textrm{TP}| + 
    |\textrm{FP}| + |\textrm{FN}|}.
\]

\paragraph{General Prediction Performance.} For $n_p$ testing
locations with true observed values $y^\mathrm{true}$, denote
$\mu^\star$ as the predictive mean and $\sigma^\star$ as the predictive
standard deviation.  Root mean squared error (RMSE) is defined as
\[
\mathrm{RMSE} = \sqrt{\frac{1}{n_p}\sum_{i = 1}^{n_p} 
    \left(\left(\mu_i^\star - y^\mathrm{true}_i\right)^2\right)}.
\]
Given a Gaussian posterior predictive distribution, continuous 
rank probability score \citep[CRPS;][]{gneiting2007strictly}
is defined as
\[
\mathrm{CRPS}\left(y^\mathrm{true} \mid \mu^\star, \sigma^\star\right) = 
\frac{1}{n_p}\sum_{i=1}^{n_p}\left[
    \sigma^\star_i \left(
    2\phi(z_i) + z_i\left(2 * \Phi(z_i) - 1 \right)
    - \frac{1}{\sqrt{\pi}}\right)\right]
\quad\textrm{for}\quad 
z_i = \frac{y_i^\mathrm{true} - \mu_i^\star}{\sigma_i^\star}
\]
where $\phi$ is the standard Gaussian pdf and $\Phi$ is the standard
Gaussian cdf.

\section{GP for plateau function}\label{app:gpplateau}

\begin{figure}[h!]
\centering
\includegraphics[width=17.5cm,trim=10 10 10 10]{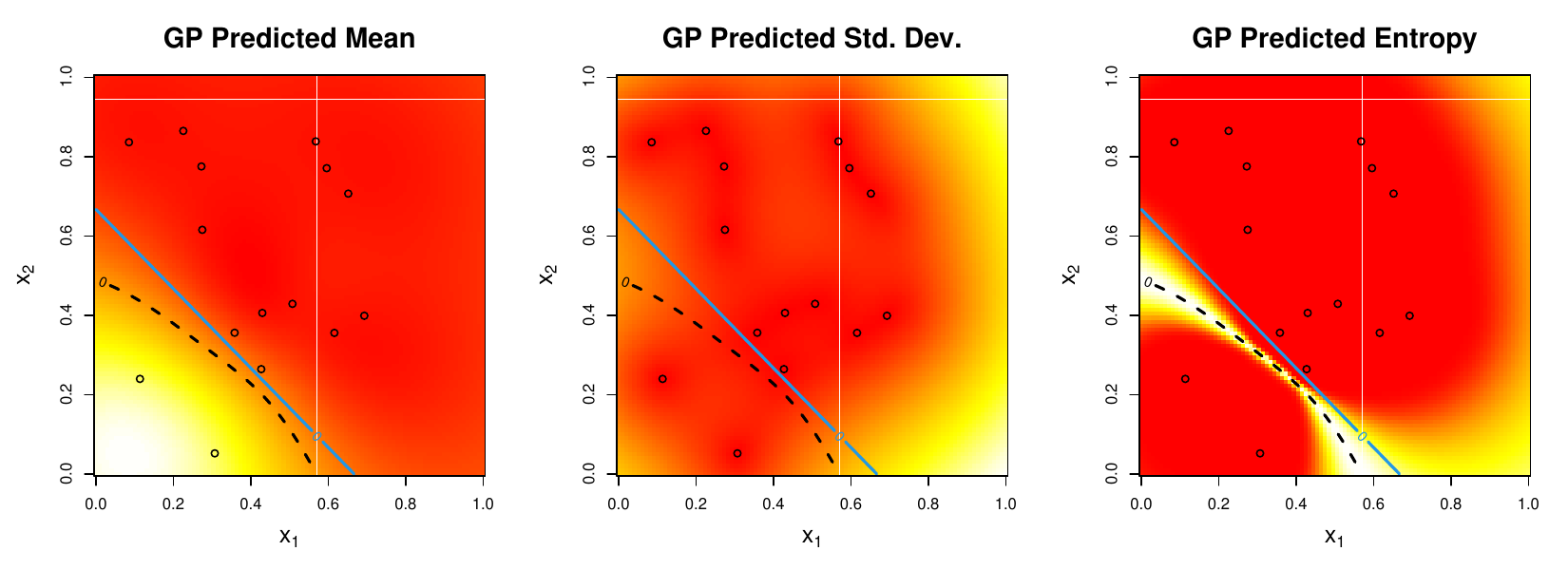}
\caption{One-layer GP predicted mean (left), standard deviation 
(center), and corresponding entropy (right) for plateau function 
(Eq.~10).  Open circles indicate training data.}
\label{fig:twoD_GP}
\end{figure}

Figure \ref{fig:twoD_GP} re-creates Figure \ref{fig:twoD} with a stationary
GP surrogate instead of a two-layer DGP.  
The GP surrogate is parameterized by
separable lengthscales sampled by Metropolis-Hastings-within-Gibbs
and was fit using the {\tt deepgp} {\sf R}-package \citep{deepgp}.  Compared
to the non-stationary DGP, the stationary GP
struggles to balance the steep drop between regimes and offers a 
much softer slope in the predictive surface that does not capture the 
plateaus as accurately (left panel).  The posterior variance of 
the GP (center panel) is solely a function of distance to 
observed training data.  Both the mean and variance are involved
in the calculation of the entropy surface (right panel).

\section{Proposed Algorithm}

The following algorithm summarizes our proposed sequential design scheme.

\medskip
\begin{algorithm}[H]
\DontPrintSemicolon
Initialize design $\{X_n, y_n\}$ and maximum budget of simulator evaluations $B$ \\
\While{$n\leq B$}{
    Train or update DGP surrogate, conditioned on observed data $\{X_n, y_n\}$ \\
    Build tricands $\mathcal{X}_N$ from $X_n$\\
    Use surrogate to evaluate entropy $H(\mathcal{X}_N)$ and uncertainty $\sigma_Y(\mathcal{X}_N)$ \\
    Identify Pareto front of uncertainty and entropy, denoted $P(\mathcal{X}_N)$ \\
    Select $x_{n+1}\sim\mathrm{Unif}[P(\mathcal{X}_N)]$ \\
    Evaluate $y_{n+1} = f(x_{n+1})$, augment $\{X_n, y_n\}$, and increment $n\leftarrow n+1$ \\
}
\end{algorithm}

\section{Computation Time}\label{app:time}

\begin{figure}[h!]
\centering
\includegraphics[width=17.5cm,trim=10 10 10 10]{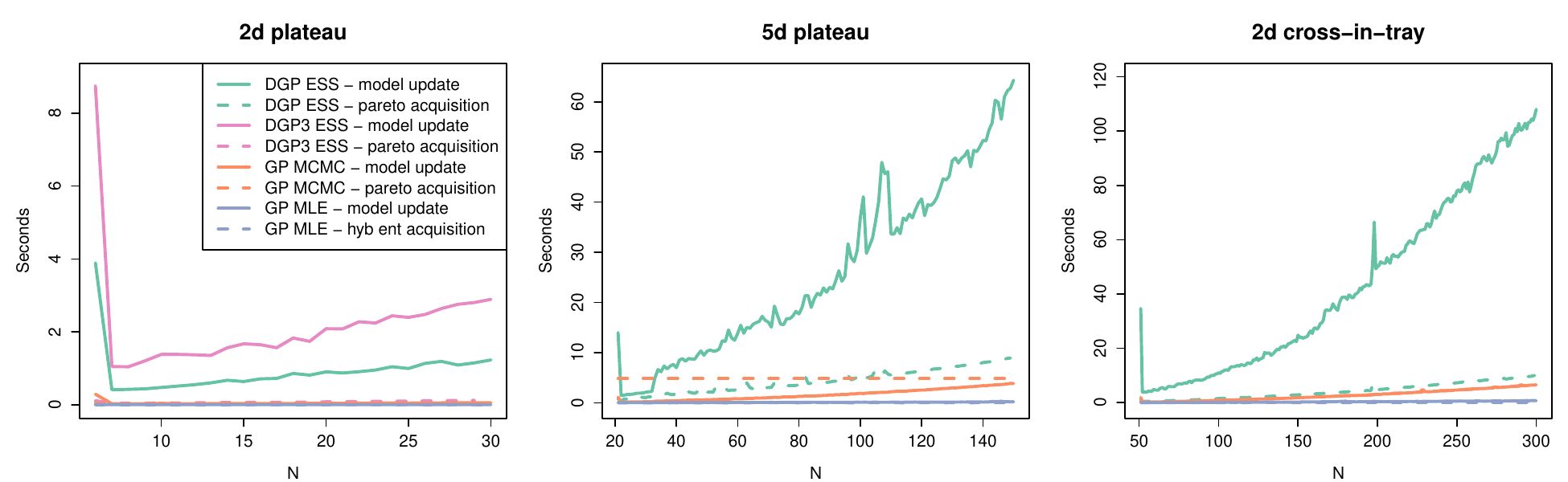}
\caption{Computation time of a single run of each synthetic 
example.  Times shown across acquisitions, separated by model training/updating
(solid) and acquisition (dashed).}
\label{fig:time}
\end{figure}

Figure \ref{fig:time} shows the computation time in seconds
for a single run of each simulated example from Section \ref{sec:synthetic}.
Recall our target application involves an expensive computer 
model which requires several minutes to obtain a single run.
Times were collected on a 13th Gen Intel Processor; for the sake of 
comparison, we restricted each method to a single CPU.
Solid lines display the time required to fit the original surrogate model
(the peak at the start) and the time required to update it after each 
acquisition.  Dashed lines show the time required to complete the 
acquisition procedure.
There are several things to note, but nothing surprising.  Both
MCMC-based models (DGP ESS and GP MCMC) are slower than the MLE-based GP.
We conducted 10,000 MCMC iterations for the initial fit followed by 1,000 
additional samples for each update.  Fewer iterations may
have sufficed.  Although all GPs experience computational costs that 
grow cubicly with $n$, the DGP with its layering of multiple GPs feels this 
impact most heavily.  For the larger data sizes, a Vecchia approximation 
could alleviate these costs \citep{katzfuss2020vecchia,sauer2022vecchia}.  
The cost of the tricands-Pareto 
acquisition procedure is most pronounced in the 5-dimensional case, in part
owing to the cost of generating the Delaunay triangulation in higher dimensions.

\end{document}